\documentclass[article,12pt]{elsarticle}

\usepackage{lineno,hyperref}

\journal{Nuclear Instruments and Methods in Physics Research Section A}
\usepackage[]{hyperref} 	
\usepackage[]{epsfig}
\usepackage{graphicx}
\usepackage{xcolor}
\usepackage{dcolumn}
\usepackage{epstopdf}
\usepackage{lineno}
\usepackage{multirow}
\usepackage{array}
\usepackage{enumitem}
\usepackage{booktabs}
\usepackage{tabularx}
\usepackage{ltablex} 
\usepackage[utf8]{inputenc} 

\newcolumntype{C}[1]{>{\centering\let\newline\\\arraybackslash\hspace{0pt}}m{#1}}

\begin{document}
\begin{frontmatter}
\title{Input Comparison of Radiogenic Neutron Estimates for Ultra-low Background Experiments\\
}

\author[1]{J.~Cooley}
\author[2]{K.J.~Palladino} 
\author[1]{H.~Qiu} 
\author[3]{M.~Selvi}  

\author[4]{S.~Scorza\corref{cor1}}
\ead{silvia.scorza@snolab.ca}
\author[5]{C.~Zhang} 
\address[1]{Department of Physics, Southern Methodist University, Dallas, TX 75275, USA}
\address[2]{Department of Physics, University of Wisconsin-Madison, Madison, WI 53706, USA}
\address[3]{INFN - Sezione di Bologna, Italy}
\address[4]{SNOLAB, Creighton Mine \#9, 1039 Regional Road 24, Sudbury, ON P3Y 1N2, Canada}
\address[5]{Department of Physics, University of South Dakota, Vermillion, USA}

\cortext[cor1]{Corresponding author}

\begin{abstract}

Ultra-low-background experiments 
address some of the most important open questions in particle physics, cosmology and astrophysics: the nature of dark matter, whether the neutrino is its own antiparticle, and does the proton decay. 

These rare event searches require well-understood and minimized backgrounds. Simulations are used to understand backgrounds caused by naturally occurring radioactivity in the rock and in every piece of shielding and detector material used in these experiments. Most important are processes like spontaneous fission and ($\alpha$,n) reactions in material close to the detectors that can produce neutrons. A comparison study between two dedicated software packages is detailed. The cross section libraries, neutron yields, and spectra from the Mei-Zhang-Hime and the SOURCES-4A codes are presented. The resultant yields and spectra are used as inputs to direct dark matter detector toy models in GEANT4, to study the impact of their differences on background estimates and fits. Although differences in neutron yield calculations up to 50\% were seen, there was no systematic difference between the Mei-Hime-Zhang and SOURCES-4A results. Neutron propagation simulations smooth differences in spectral shape and yield, and both tools were found to meet the broad requirements of the low-background community.

\end{abstract}
\end{frontmatter}
%
\section{Introduction}\label{sec:intro}

Reduction of radiogenic backgrounds is one of the most important factors for rare event search experiments including searches for dark matter, neutrinoless double beta decay, proton decay, and detection of solar and supernovae neutrinos.  In the dark matter community, these radiogenic backgrounds are generically classified into two types:  electron recoil and nuclear recoil backgrounds.  Electron recoil backgrounds result from the interactions of gamma rays, electrons or beta particles interacting with the electrons in the detector's target medium while nuclear recoil backgrounds result from neutrons interacting with the nucleus in the detector's target medium. 
  In dark matter experiments, the nuclear recoil background is of particular concern because a nuclear recoil from a neutron can be indistinguishable from a nuclear recoil from a WIMP \cite{Jungman:1995df}.
  
Neutrons are produced by spontaneous fission, ($\alpha$, n) and muon-induced interactions.  
Concerns of ($\alpha$, n) interactions have been a constant from early low background experiments~\cite{Heaton:1989vbt, Formaggio:2004ge} to the current and planned direct dark matter detectors~\cite{Akerib:2014rda, Cai:2011zzc, Aprile:2015uzo, Akerib:2015cja, Agnese:2016cpb}. Muon-induced neutrons can be reduced by operation of the detectors deep underground and by placing passive shielding around the detector.  In addition, muon-induced neutrons can often be recognized by identifying the parent muon in a muon veto.  More difficult to deal with are neutrons resulting from ($\alpha$-n) interactions and spontaneous fission from $^{238}$U, $^{235}$U, and $^{232}$Th present in the materials used to construct the shielding and the detectors themselves.  
These backgrounds have induced the development of external neutron vetoes~\cite{Wright:2010bu, Calkins:2015hya, Akerib:2015cja}, and the development of robust screening and assaying programs \cite{Loach:2016fsk} for materials used in the construction of such experiments.

Two codes available for use in such calculations, SOURCES-4A~\cite{Wilson2009608} and a code developed by Mei, Zhang and Hime~\cite{Mei:2008ir} that we will refer to as the USD webtool, are evaluated here, extending the work of ~\cite{Selvi:2013nru}. 
The latter is made available online at http://neutronyield.usd.edu. A user can enter information such as the decay chain and the material details to consider, but can not access nor modify the code. The SOURCES-4C code is available through the Radiation Safety Information Computational Center (RSICC) at Oak Ridge National Laboratory (US) (http://rsicc.ornl.gov). A modified  version of the original SOURCES-4A version has been obtained via email exchange with Dr. V. Kudryavtsev at University of Sheffield~\cite{VK} and was used in this study. Modifications of this version are detailed in the next section.

Differences exist in how each code calculates neutron spectra. The USD code allows calculation for the whole $^{238}$U, $^{232}$Th decay chains and Sm radioactive source, whereas SOURCES-4A  allows calculations of sub-chains, according to the input parameter that the user wants to run. That said, SOURCES-4A  requires the installation of the code and construction of an input tape in order to perform calculations which requires a familiarity with coding that is not required to use the web interface provided by the USD webtool.

\section{Neutron Yield and Spectra}\label{sec:neutronprod}

Measurements of neutron spectra strongly depend on the material and are not straightforward to make since neutrons are neutral particles.  As such, their calculations are critical for low background experiments. The total neutron yield indicates the number of neutrons which are produced or had entered the target  whereas                                                                                                                                                                                                                                                                                                                                                                                                                                                                                                                                                                                                                                                                                                                                                                                                                                                                                                                                                               the neutron energy spectrum  determines the background events expected in the energy range of interest. Both are needed in order to carry out a complete and reliable neutron background simulation.
Here we evaluate neutron yields and spectra performed via two different codes: a modified version of SOURCES-4A and the USD webtool.
The SOURCES-4A code has been modified to extend the cross section for ($\alpha$,n) up to 10~MeV, based on experimental data, whenever possible, and calculations performed via the EMPIRE code~\cite{EMPIRE} by the group at the University of Sheffield~\cite{Tomasello:2006sm}. 
The SOURCES-4A and USD webtool codes calculate neutron yields and spectra from ($\alpha$, n) reactions due to the decay of radionuclides. 
Radiogenic neutrons result from the decay of the intrinsic contamination of materials surrounding the detectors with  $^{232}$Th, $^{238}$U and $^{235}$U. 

A validation of the SOURCES-4A and USD webtool codes has been carried out. We have considered a straightforward alpha-beam problem: a 5.5~MeV alpha particle incident on a magnesium slab (data from Jacobs, 1983)~\cite{Jacobs:1983}.
The  integrated neutron yield results in a good agreement: 11.6$\cdot 10^{-7}$~n$\cdot$s$^{-1}$$\cdot$cm$^{-3}$ and $11.7\cdot 10^{-7}$~n$\cdot$s$^{-1}$$\cdot$cm$^{-3}$ for the SOURCES-4A and USD webtool codes, respectively.
%

A comparison study between the SOURCES-4A and USD webtool codes was performed considering both the $^{232}$Th and the $^{238}$U decay chains in secular equilibrium, although a possibility of disequilibrium can be taken into account.  Due to different migration, the long-lived isotopes, $^{226}$Ra, $^{222}$Rn, $^{210}$Po, $^{228}$Ra, $^{228}$Th and their associated decay daughters could be calculated separately, as shown in Table ~\ref{tab:contributions}. For both chains, the majority of neutrons are produced by the $\alpha$ generated in the second part of the chains. 

The calculation of the neutron spectra requires as inputs the cross-sections of ($\alpha$, n) reactions, the probabilities of nuclear transition to different excited states (branching ratios) and the alpha emission lines from the radionuclides.
Both codes consider a thick target: calculation of neutron yields and spectra are carried out under the assumption that the size of  radioactive sample exceeds significantly the range of the alpha particle.
The energy bin size of the ($\alpha$, n) calculation is fixed to be 0.1~MeV in USD webtool code whereas it is user dependent in SOURCES-4A.

Table~\ref{tab:alphaline} lists the $\alpha$-lines resulting from the $^{238}$U and $^{232}$Th decay chains that are included in the SOURCES-4A and USD webtool codes. The USD webtool code is missing the $\alpha$-lines from the $^{222}$Ra isotope in the $^{238}$U decay chain that is considered by the SOURCES-4A library. Overall, the branching ratio and the energy lines are in good agreement between the two codes.

Cross-sections and branching ratios are required for neutron yield and spectra calculations as well.  
The USD webtool code uses TENDL 2012~\cite{TENDL12} to provide  ($\alpha$, n) nuclear cross-sections. TENDL is a validated nuclear data library which provides the output of the TALYS~\cite{TALYS} nuclear model code system; the SOURCES-4A cross section input libraries come from EMPIRE-2.19~\cite{EMPIRE} calculations and, for some isotopes, a combination of data measurements and EMPIRE-2.19 calculations. EMPIRE is the code recommended by International Atomic Energy Agency (IAEA). However, neither EMPIRE nor TALYS can properly calculate all resonance behavior that has been experimentally observed.

\begin{figure}[!htb]
\begin{center}
\begin{tabular}{ll}\\
 \includegraphics[width=5.cm]{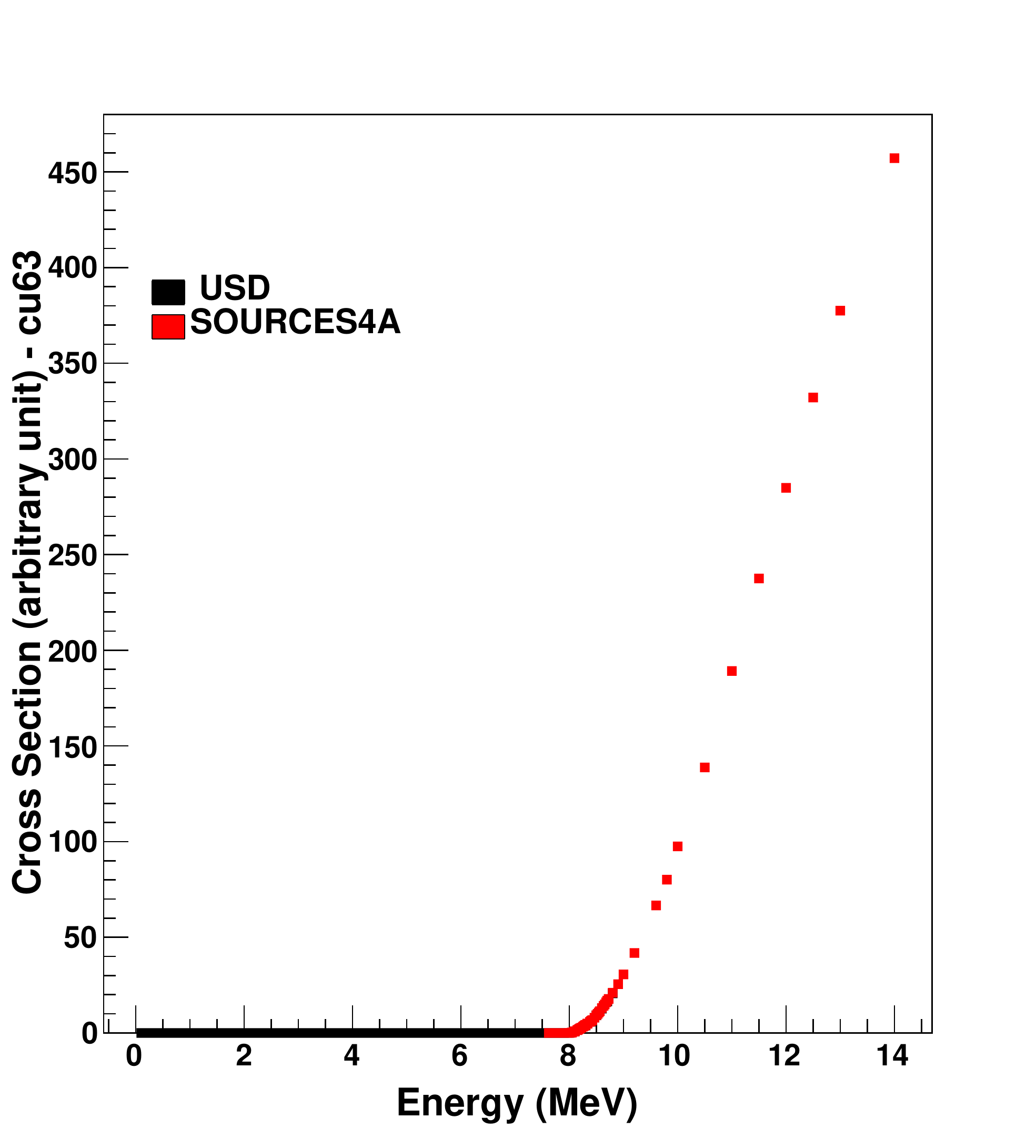} & \includegraphics[width=5.cm]{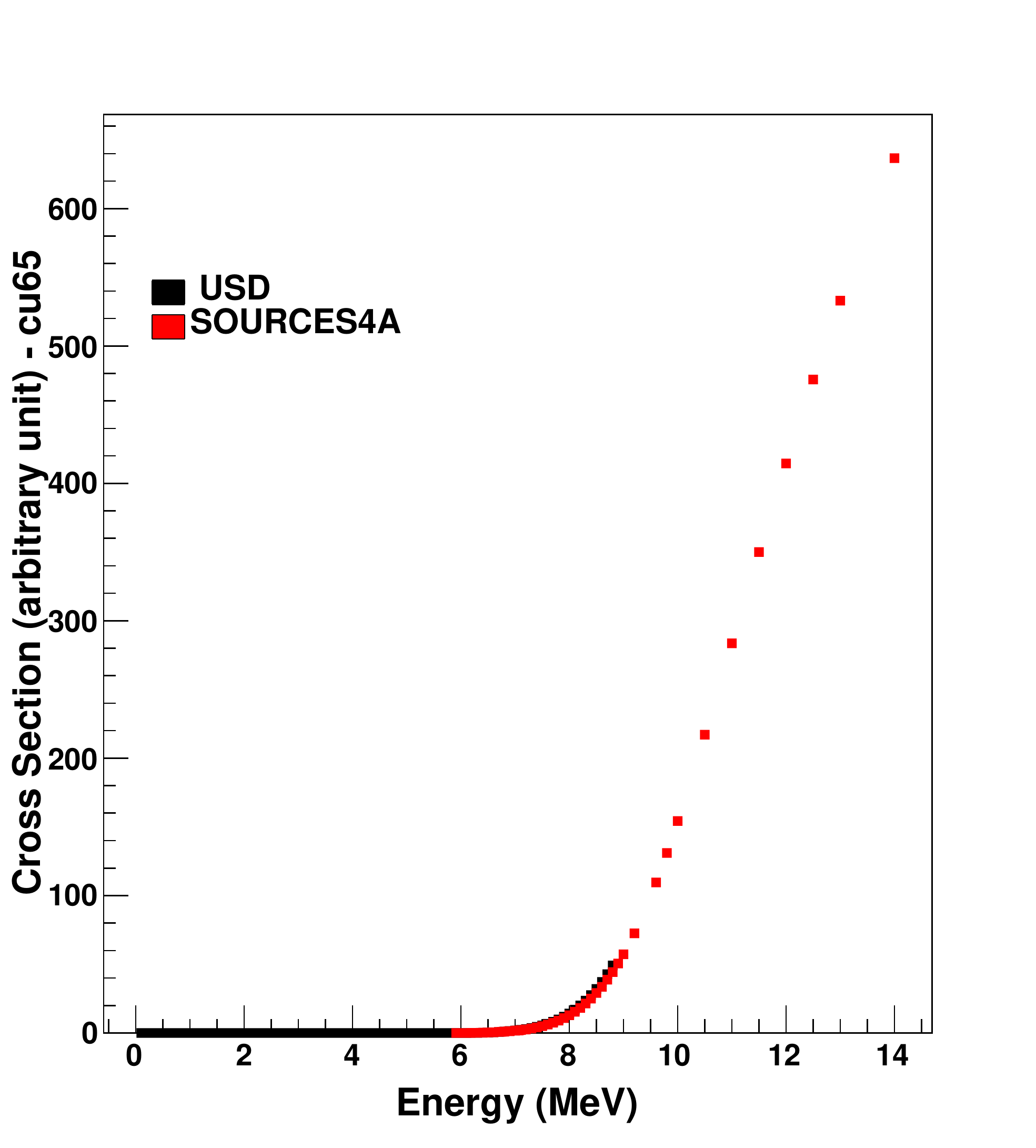} \\
 \includegraphics[width=5.cm]{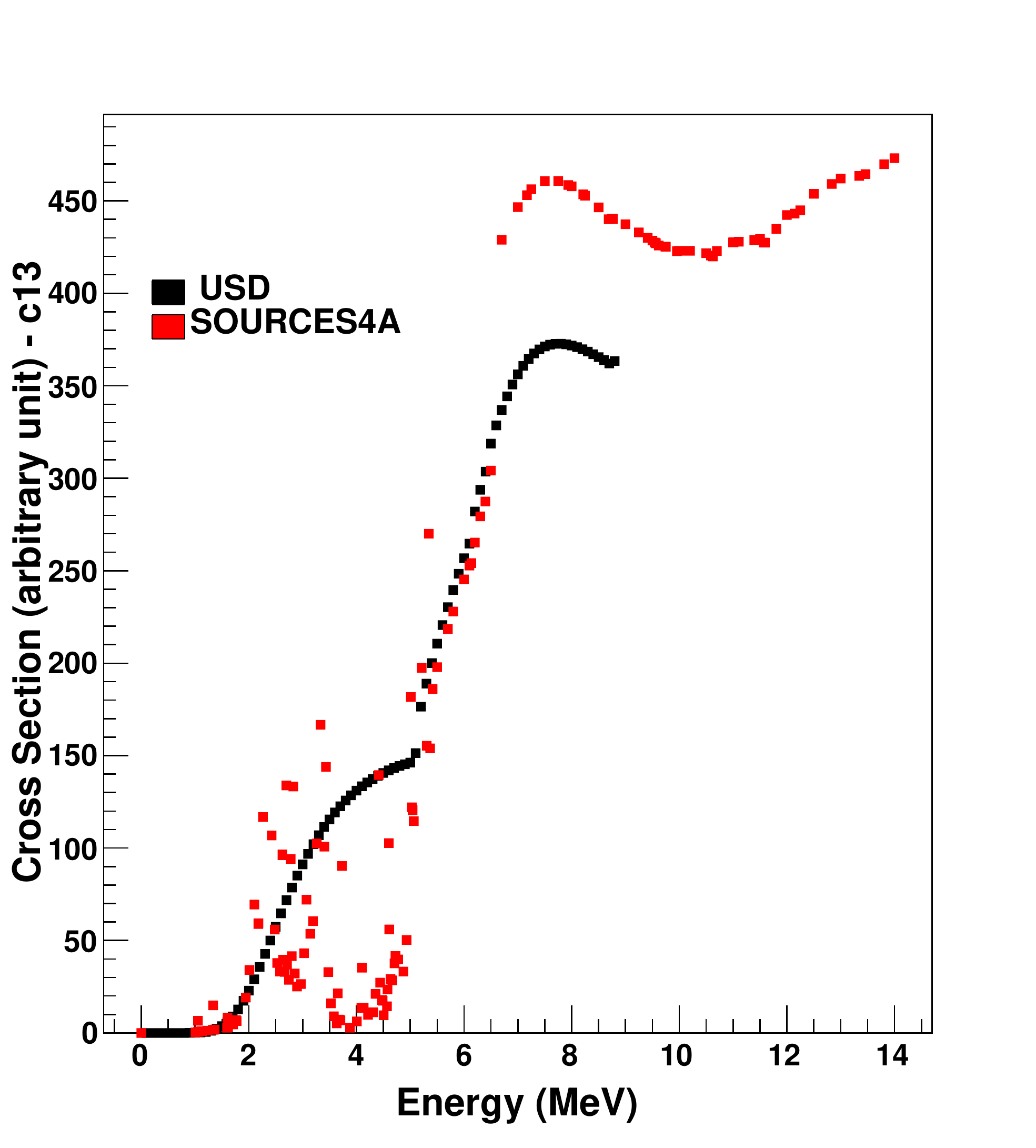} & \includegraphics[width=5.cm]{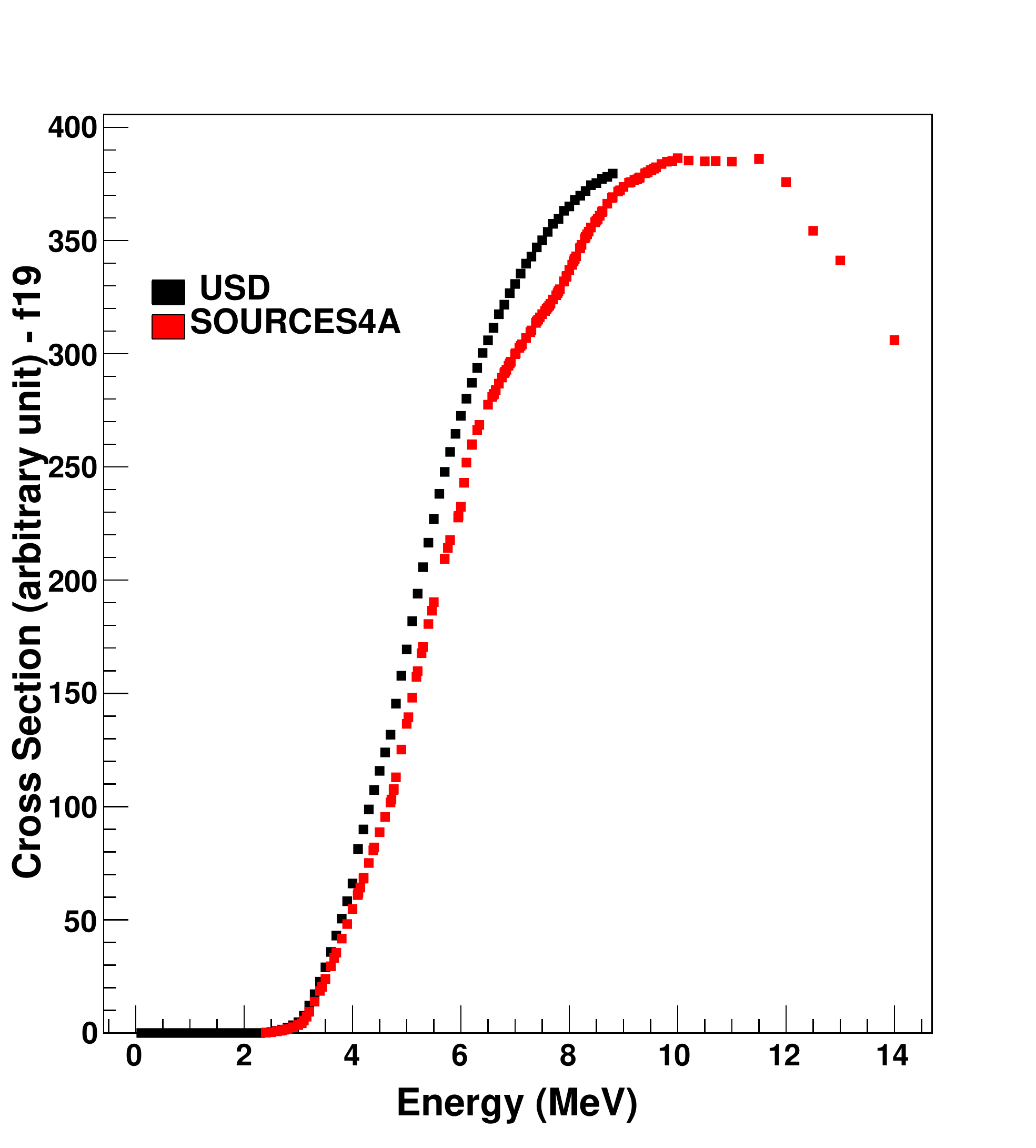} \\
 \includegraphics[width=5.cm]{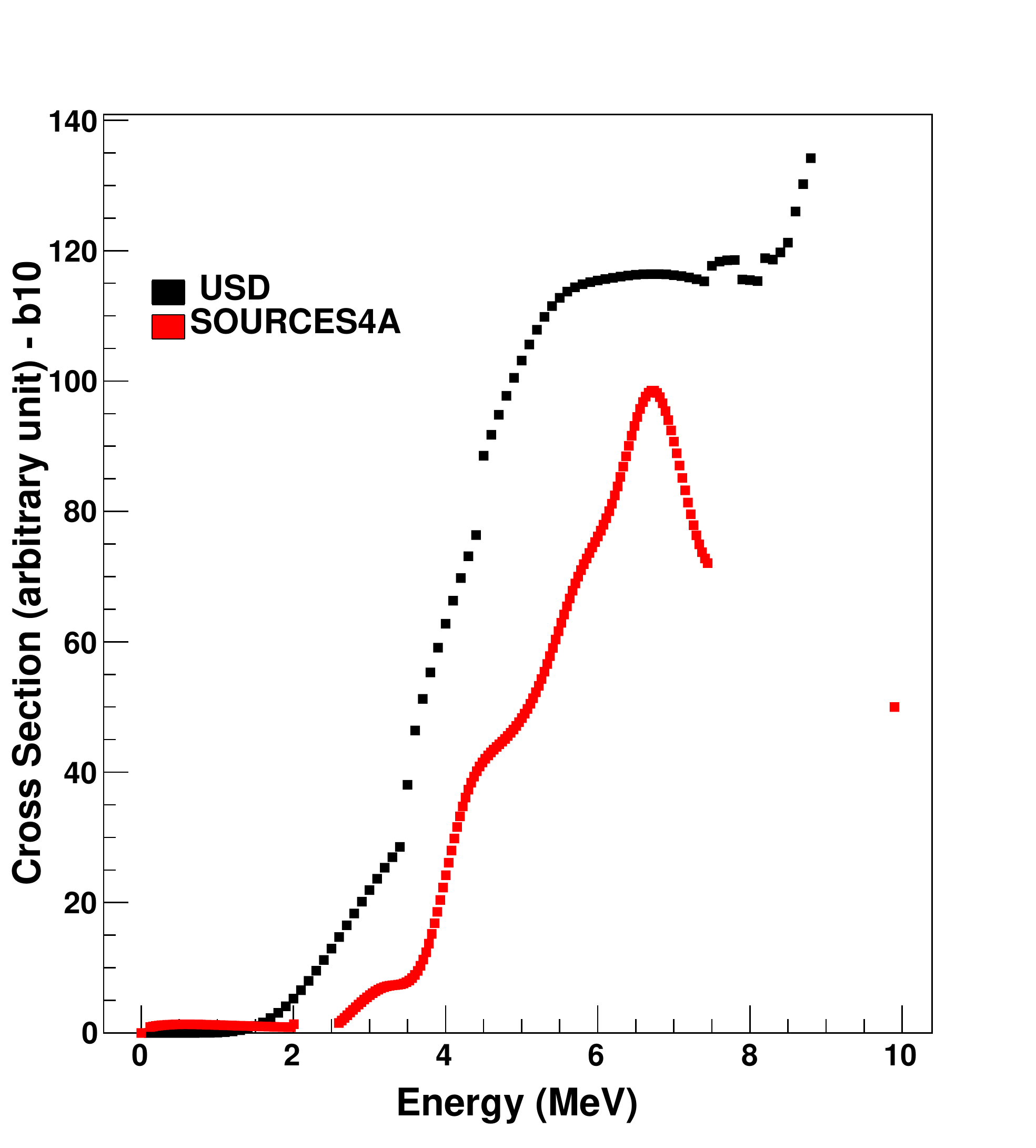} & \includegraphics[width=5.cm]{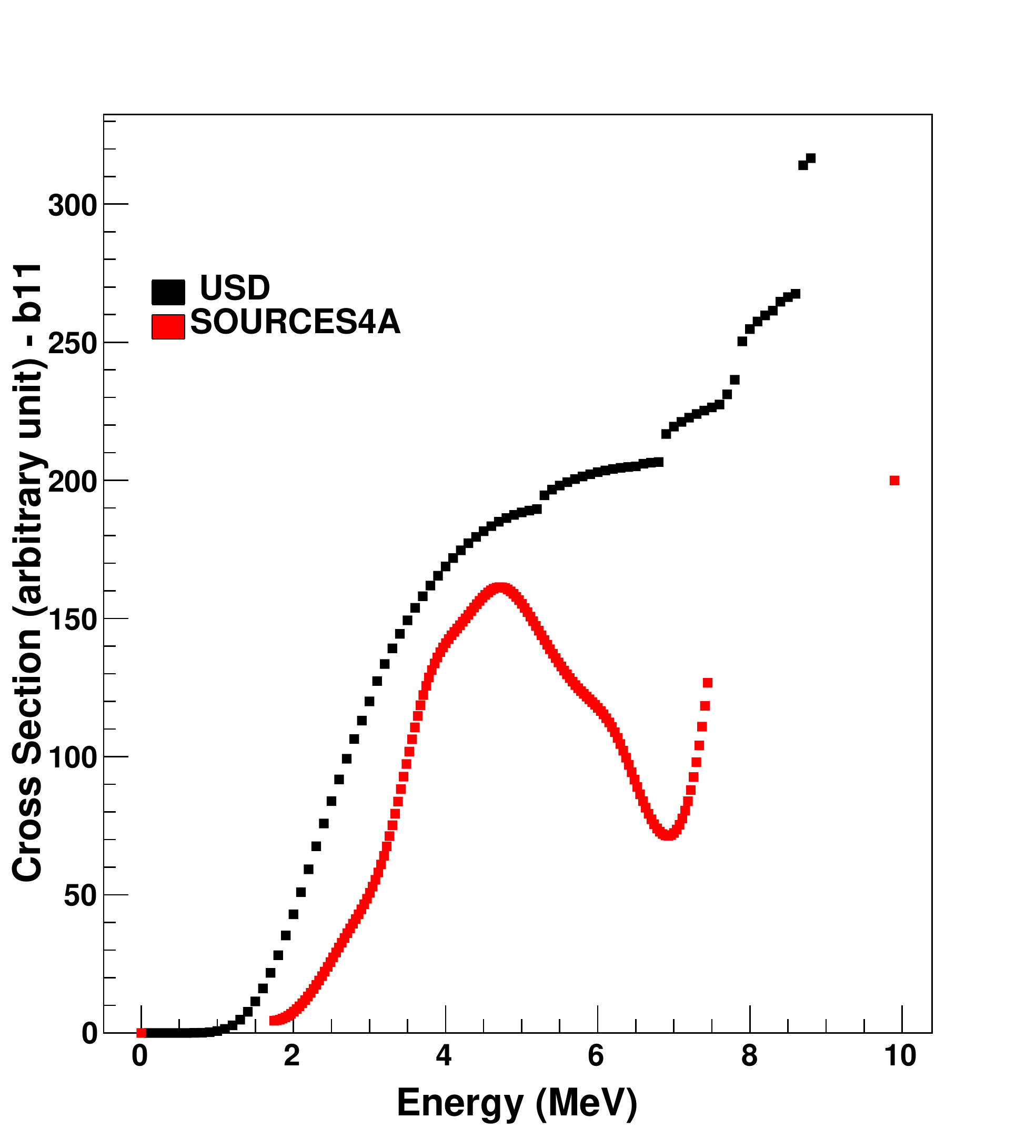} \\

\end{tabular}
\end{center}
\caption{Input ($\alpha$, n) cross-section for the target isotopes involved in radiogenic neutron calculations for copper and polyethylene. Red markers represent the SOURCES-4A code inputs, whereas the black markers represent the USD webtool code inputs. From left to right, top to bottom: $^{63}$Cu, $^{65}$Cu, $^{13}$C, $^{19}$F, $^{10}$B and $^{11}$B. 
\label{fig:crosssection}}
\end{figure}

An extensive comparison between the cross-section libraries used in the SOURCES-4A and USD webtool codes has been carried out. We considered the target nuclides present in the materials with the greatest contribution to the radiogenic neutron background. Specifically, we examined materials that commonly compose the shielding scheme and the internal detector components, such as stainless steel, copper, titanium, borosilicate glass (PMTs glass), and PTFE, which becomes important when the external neutron flux is attenuated by the shielding. The cross section spectrum results in a good agreement for most of isotopes in both code libraries. However, for some, such as $^{13}$C and $^{10}$B and $^{11}$B, the cross sections show discrepancies, see Fig.~\ref{fig:crosssection}. 

\begin{figure*}[!htb]
\centering
\begin{tabular}{ll}\\
\includegraphics[width=5.5cm]{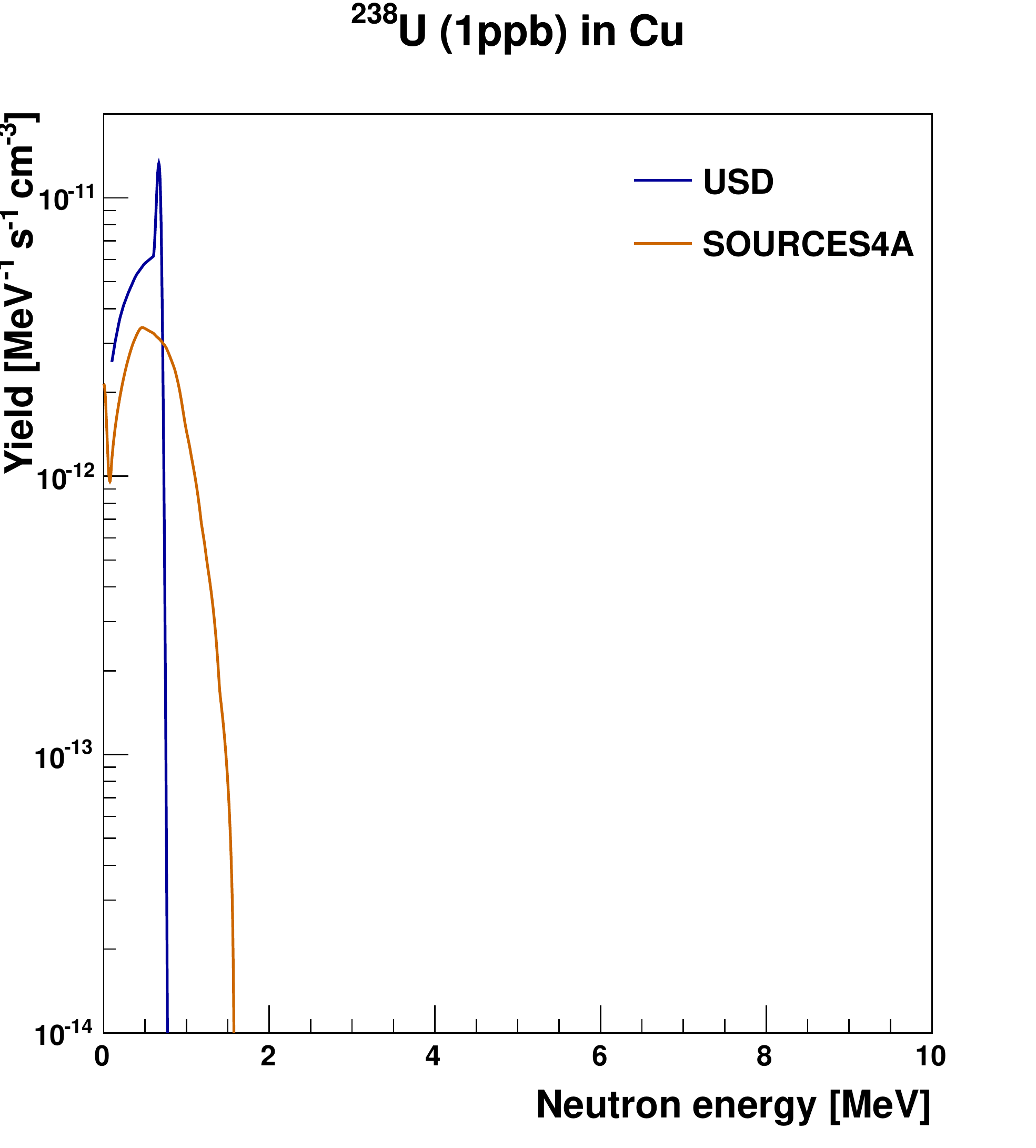} & \includegraphics[width=5.5cm]{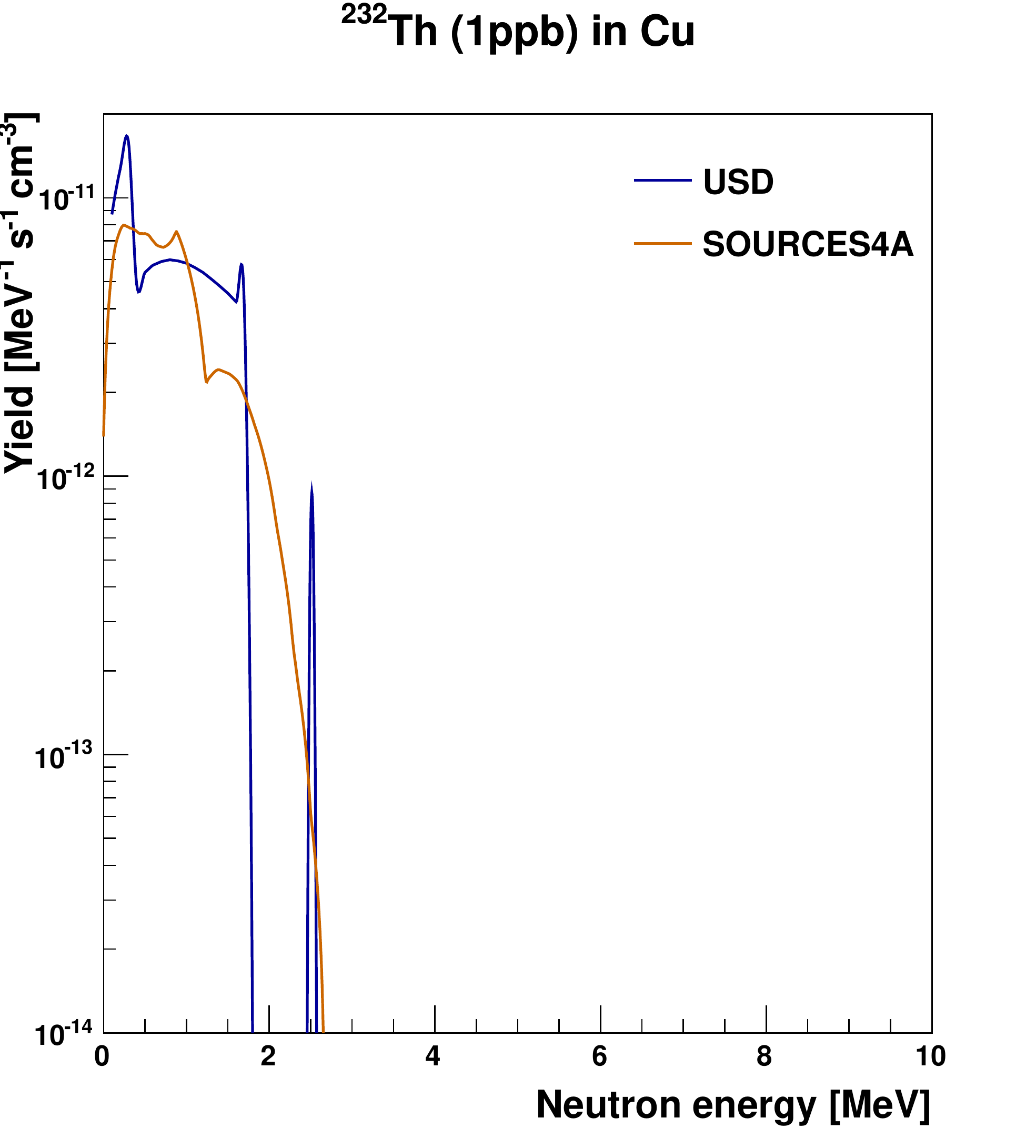}\\
\includegraphics[width=5.5cm]{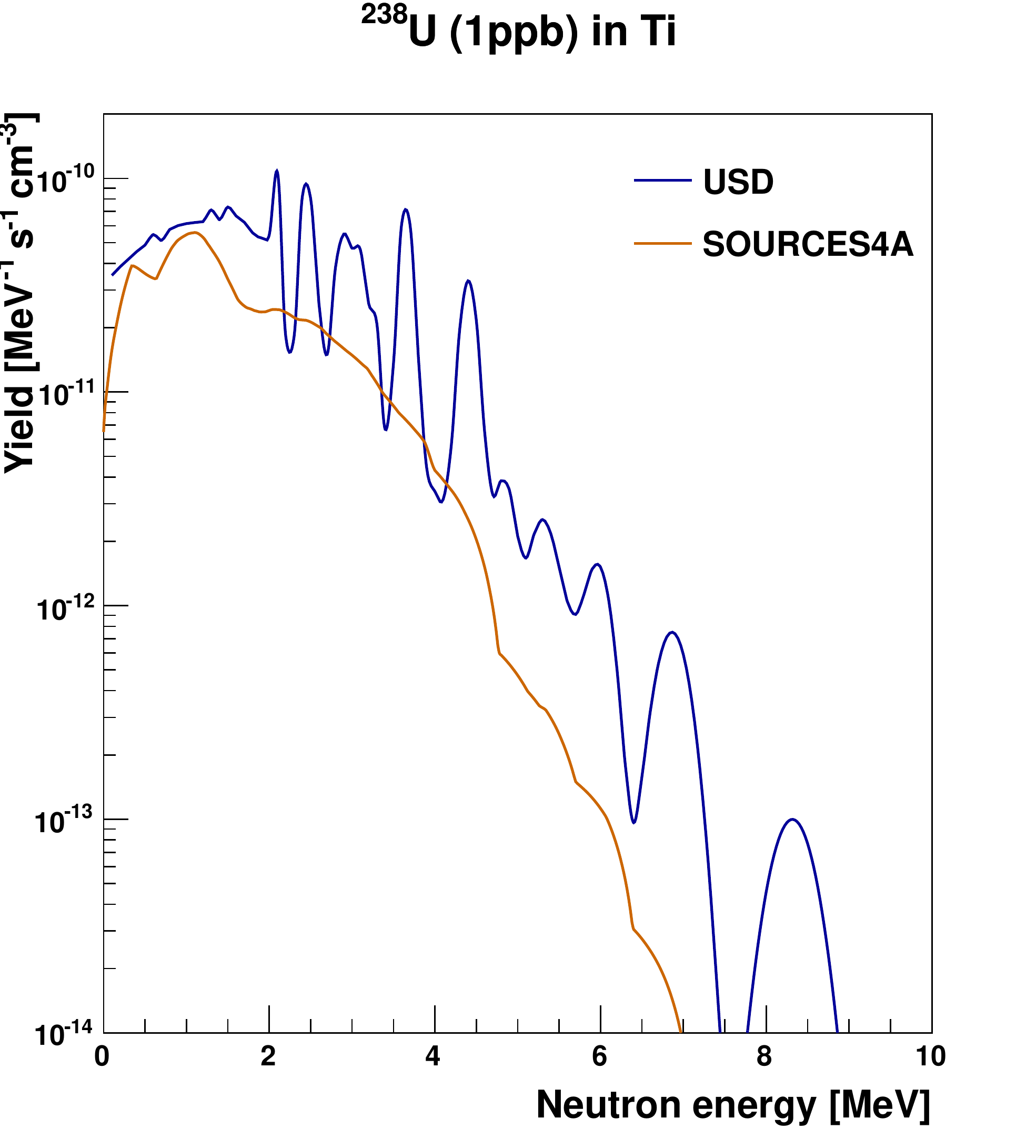} & \includegraphics[width=5.5cm]{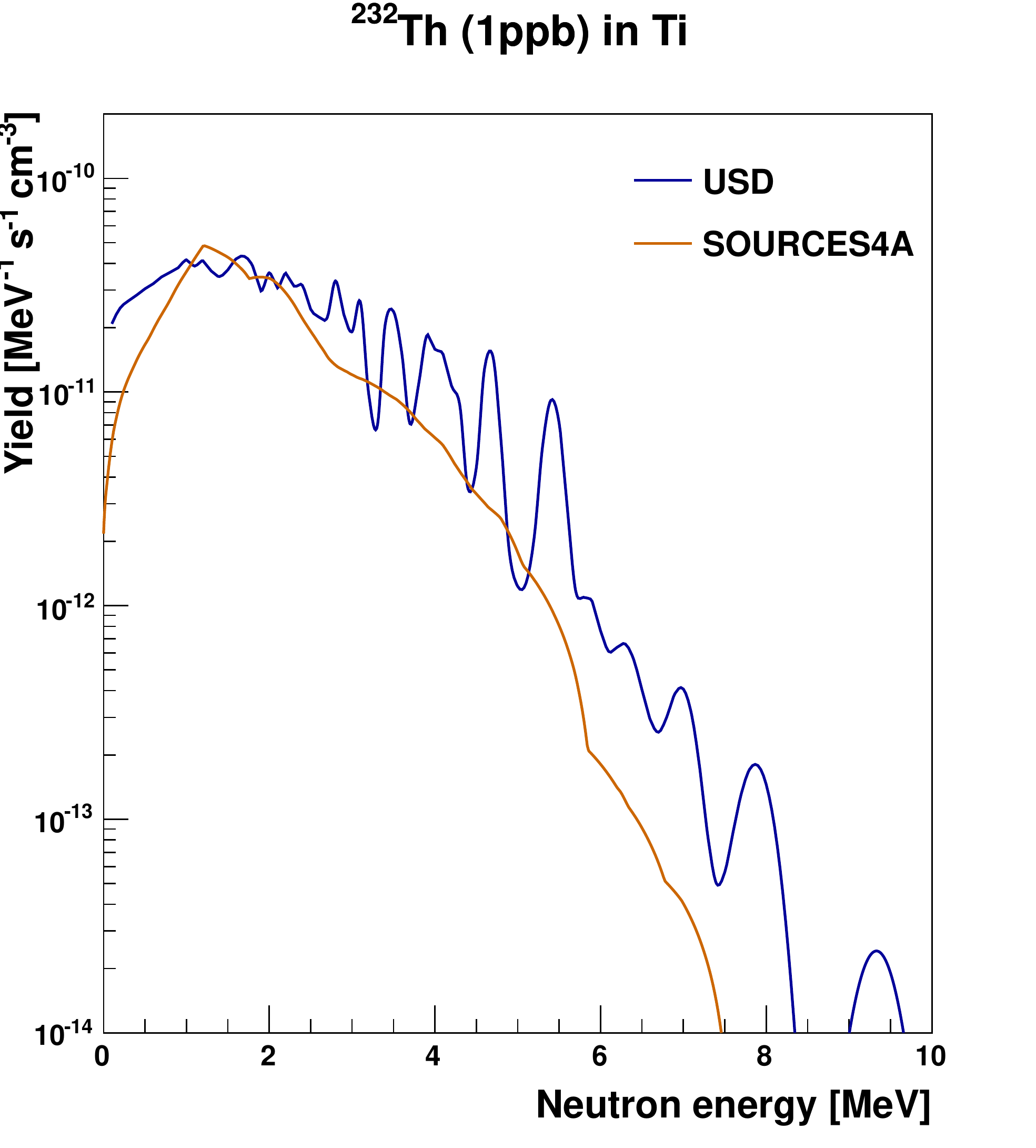} \\
\includegraphics[width=5.5cm]{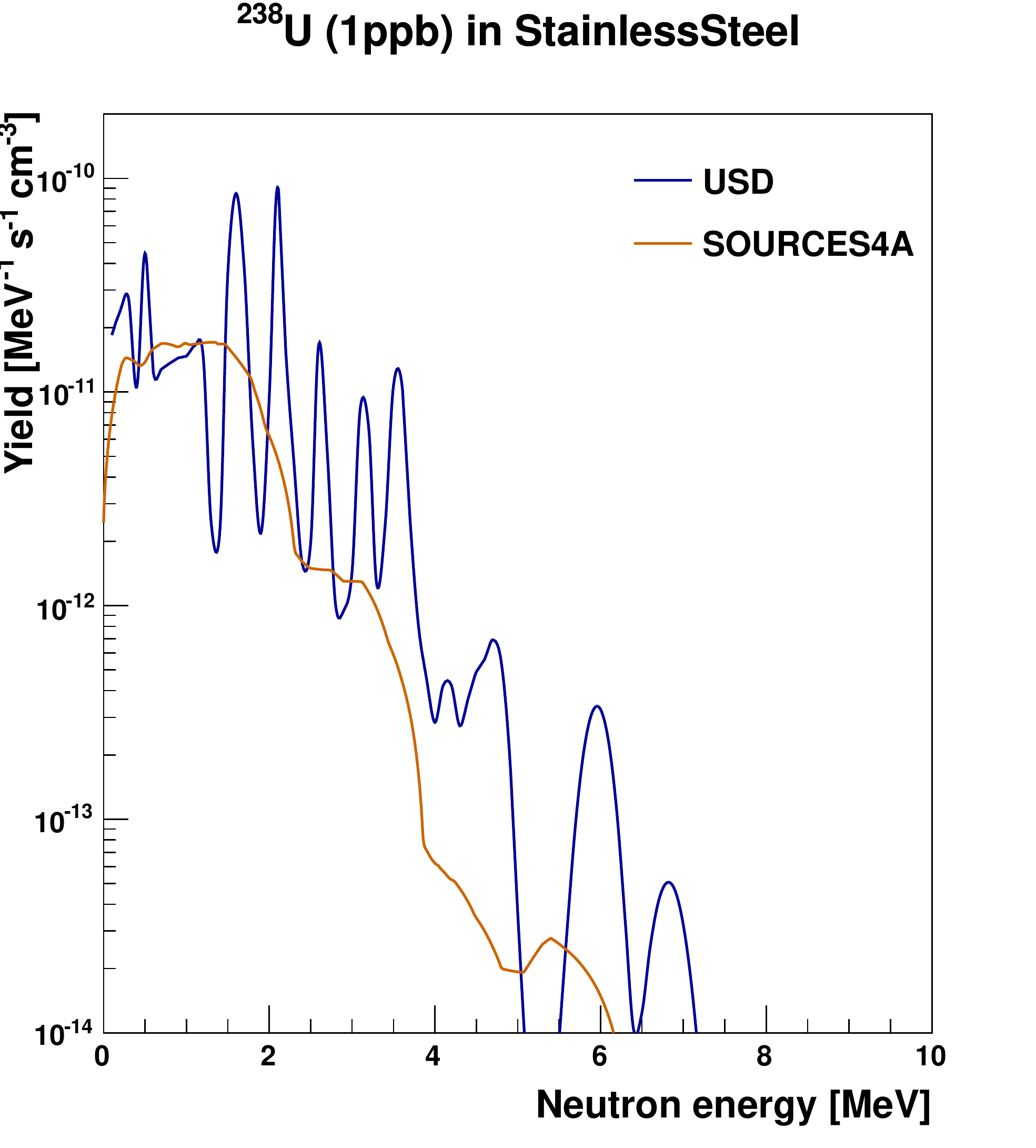} & \includegraphics[width=5.5cm]{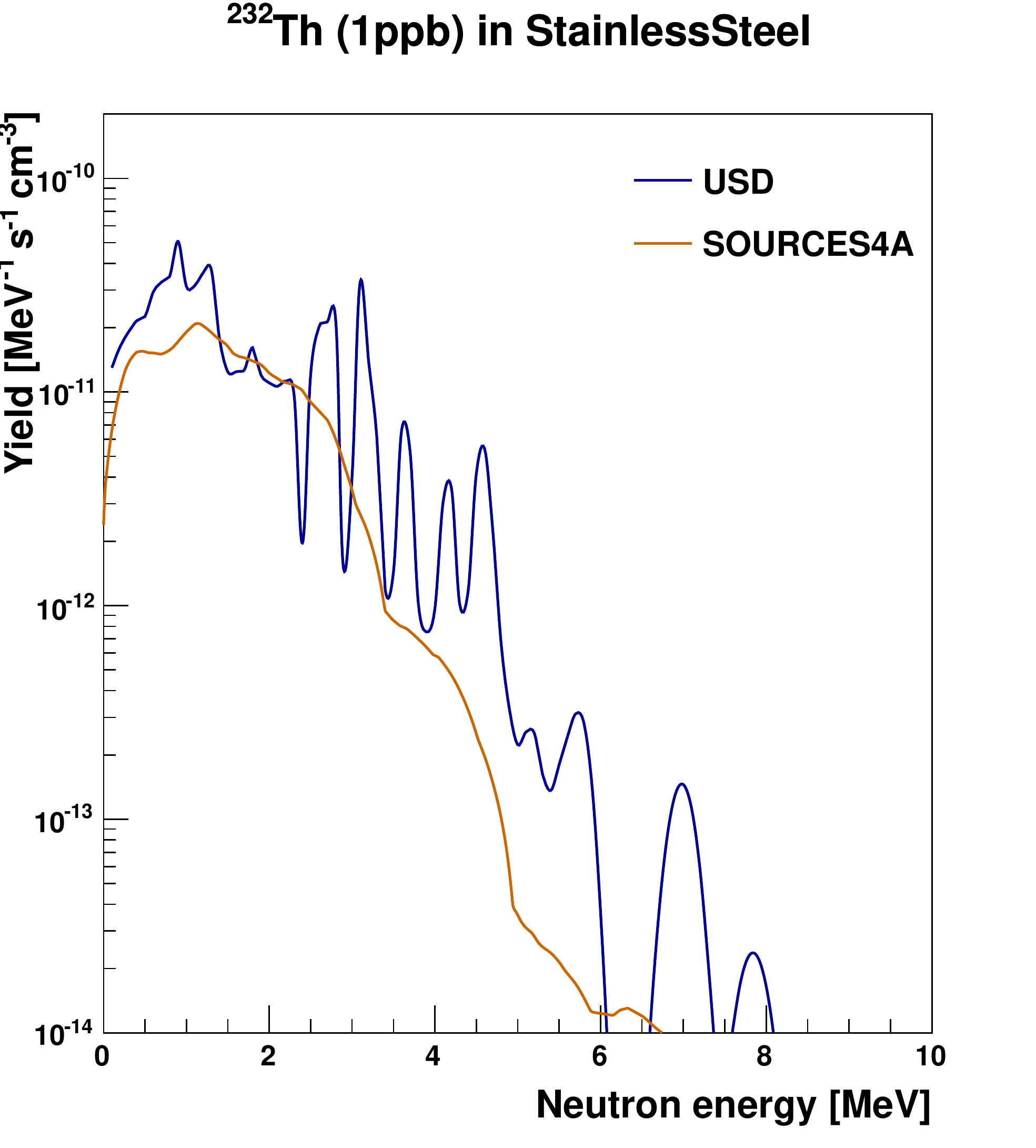}\\
\end{tabular}
\caption{Radiogenic neutron spectra (n$\cdot$MeV$^{-1}$$\cdot$s$^{-1}$$\cdot$cm$^{-3}$) calculated for 1~ppb $^{238}$U and $^{232}$Th decay chains, left and right panels, respectively. The ($\alpha$, n) reaction contribution is shown in orange for the SOURCES-4A  code and in blue for the USD webtool code.  From top to bottom materials are: copper, titanium and stainless steel.
\label{fig:allradspectra1}}
\end{figure*}

\begin{figure*}[!htb]
\centering
\begin{tabular}{ll}\\
\includegraphics[width=5.5cm]{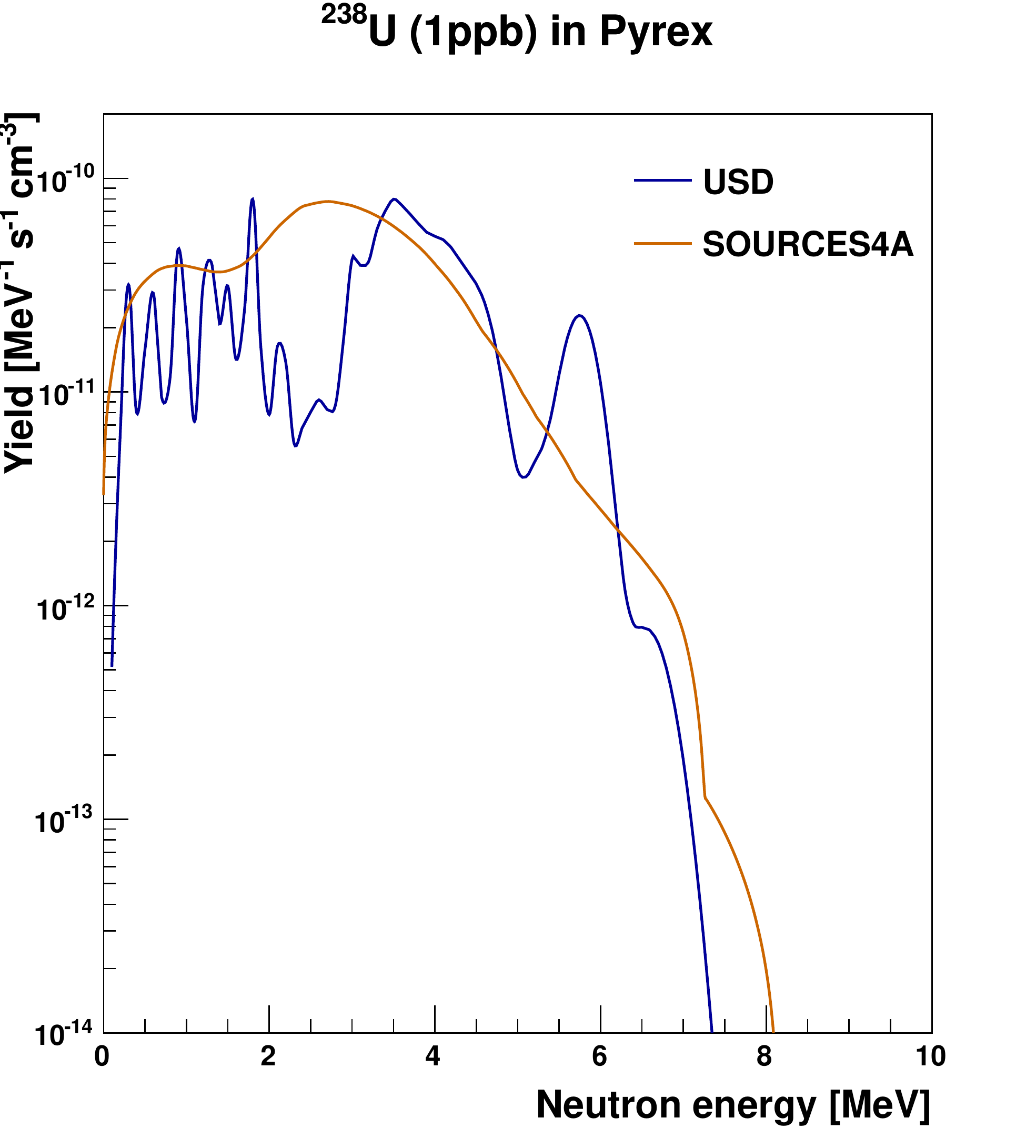} & \includegraphics[width=5.5cm]{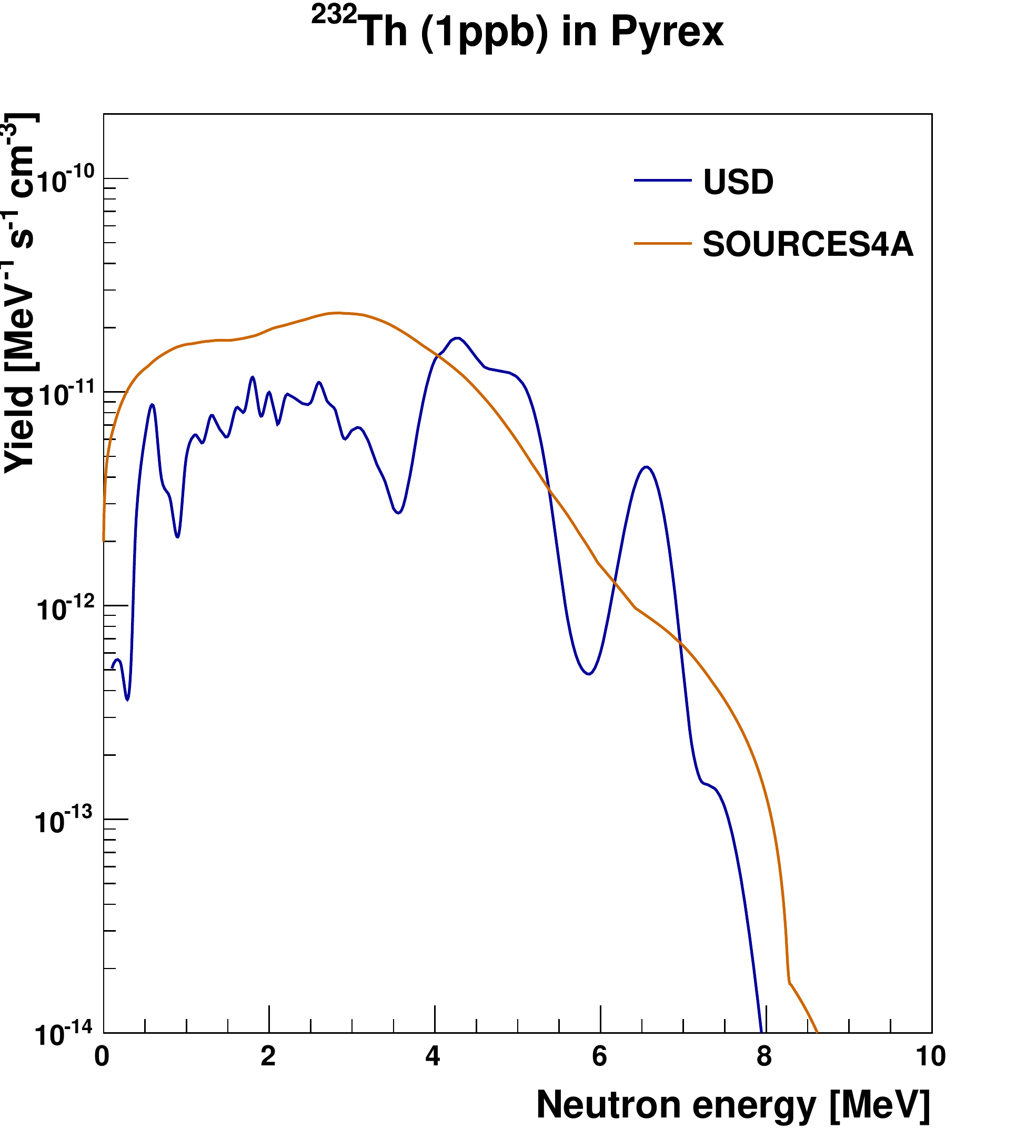} \\
\includegraphics[width=5.5cm]{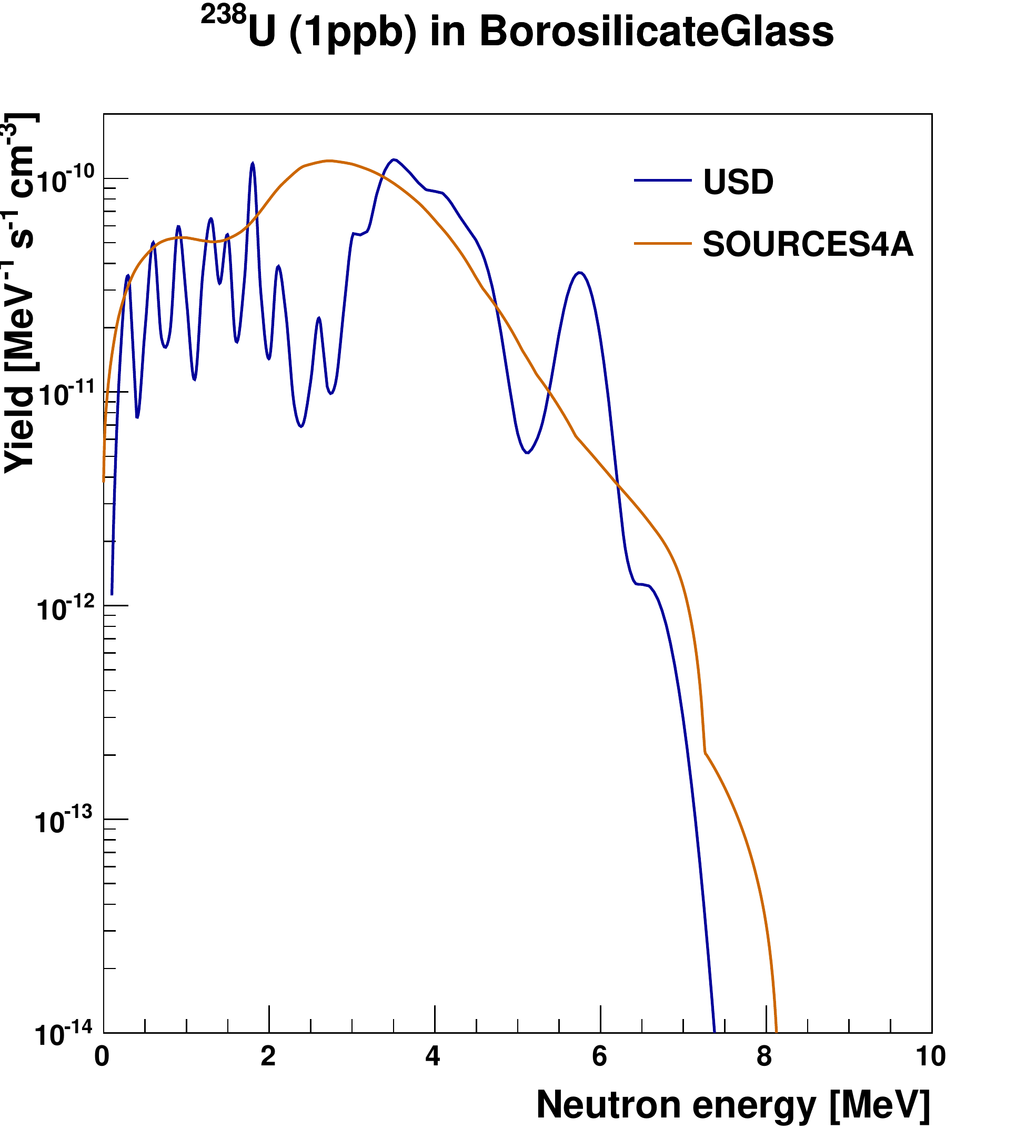} & \includegraphics[width=5.5cm]{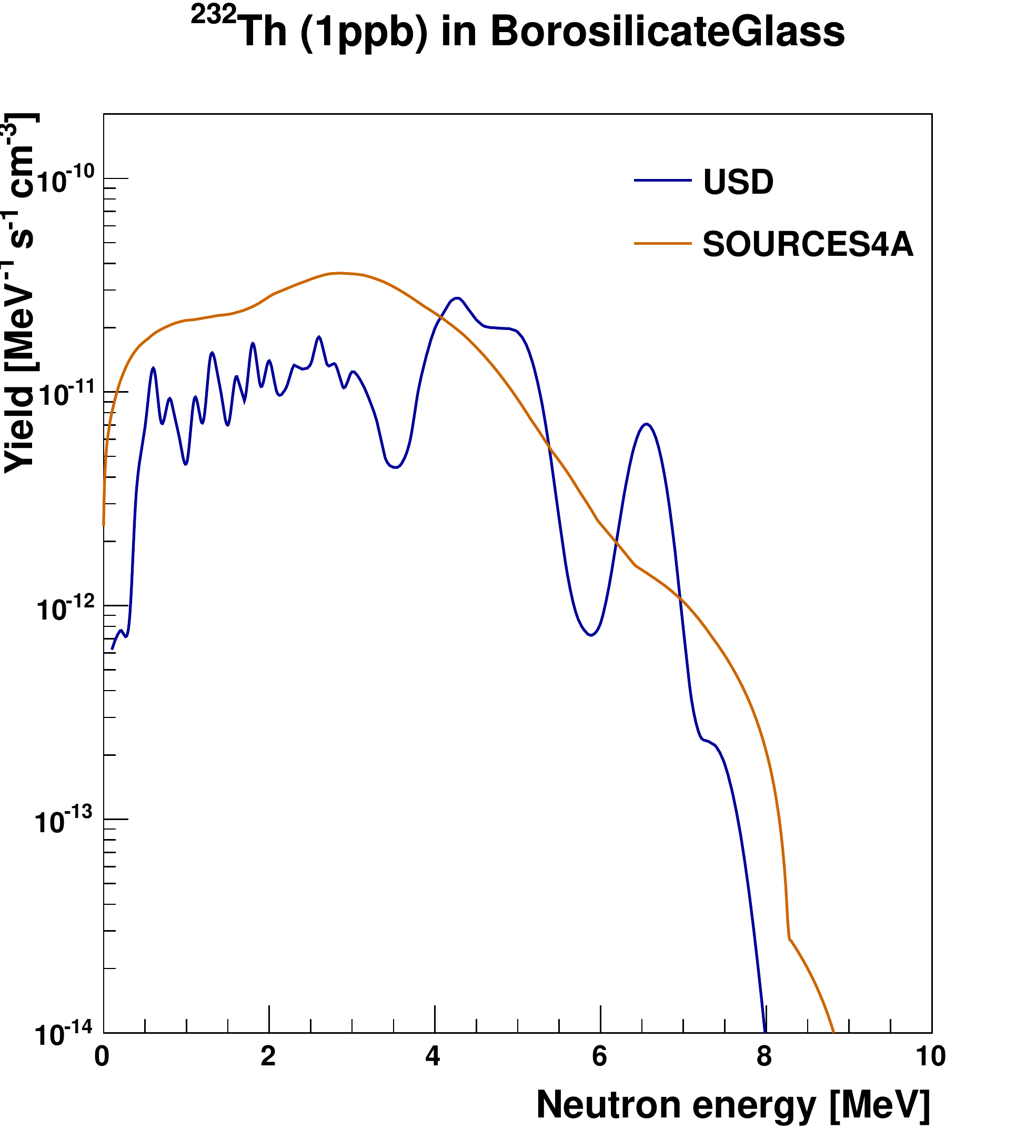}\\
\end{tabular}
\caption{Radiogenic neutron spectra (n$\cdot$MeV$^{-1}$$\cdot$s$^{-1}$$\cdot$cm$^{-3}$) calculated for 1~ppb $^{238}$U and $^{232}$Th decay chains, left and right panels, respectively. The ($\alpha$, n) reaction contribution is shown in orange for the SOURCES-4A  code and in blue for the USD webtool code.  From top to bottom materials are:  pyrex and  borosilicate glass.
\label{fig:allradspectra2}}
\end{figure*}

\begin{figure*}[!htb]
\centering
\begin{tabular}{ll}\\
\includegraphics[width=5.5cm]{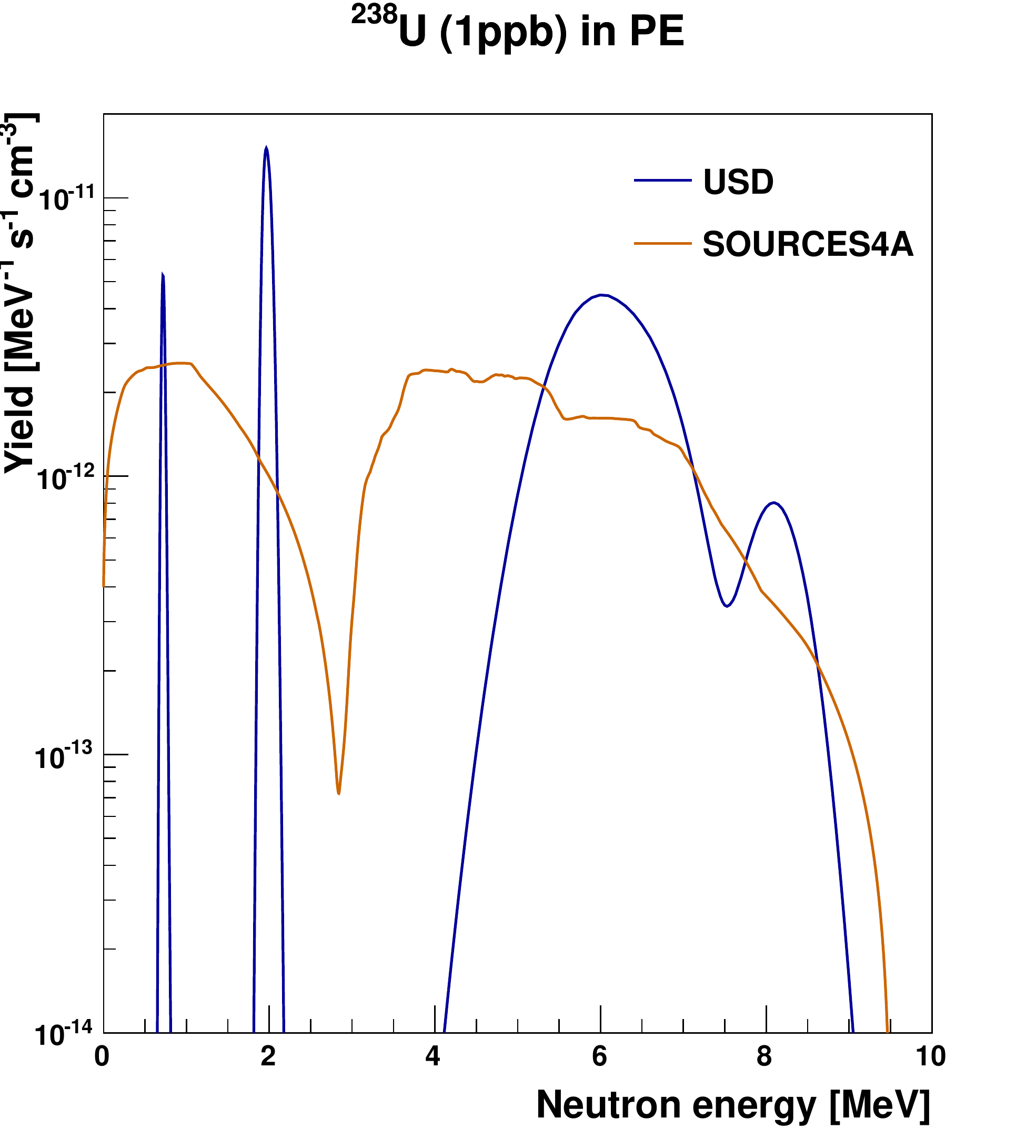} & \includegraphics[width=5.5cm]{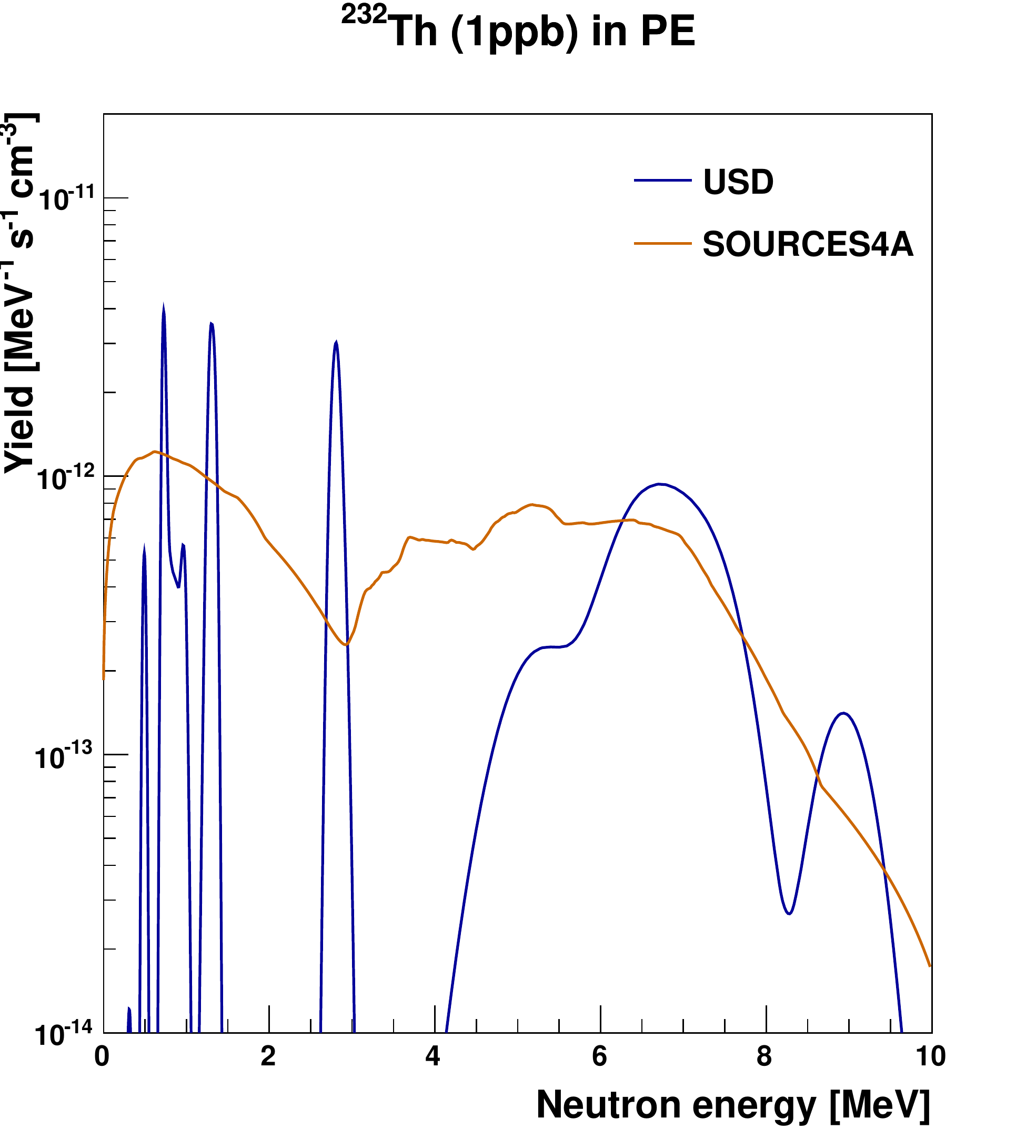} \\
\includegraphics[width=5.5cm]{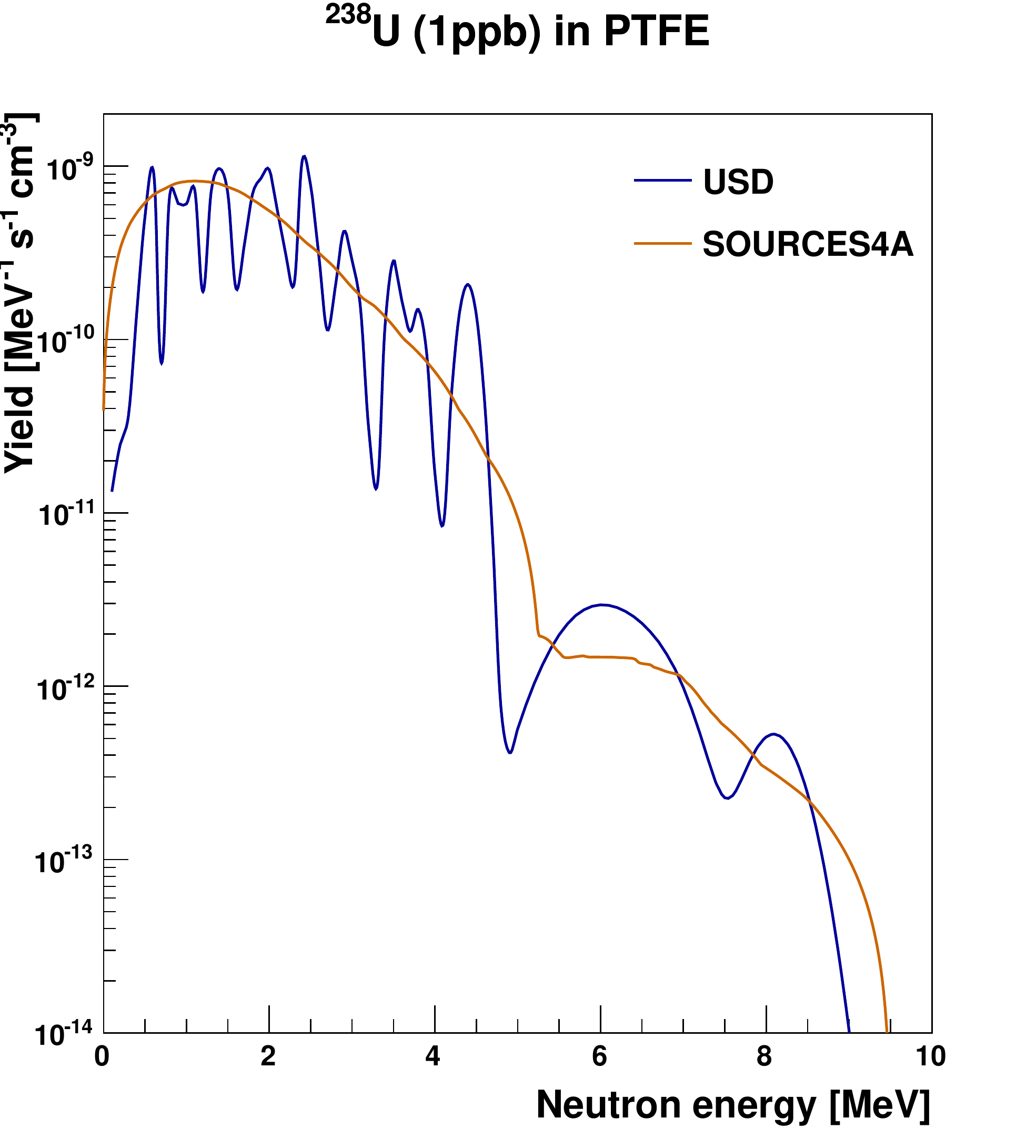} & \includegraphics[width=5.5cm]{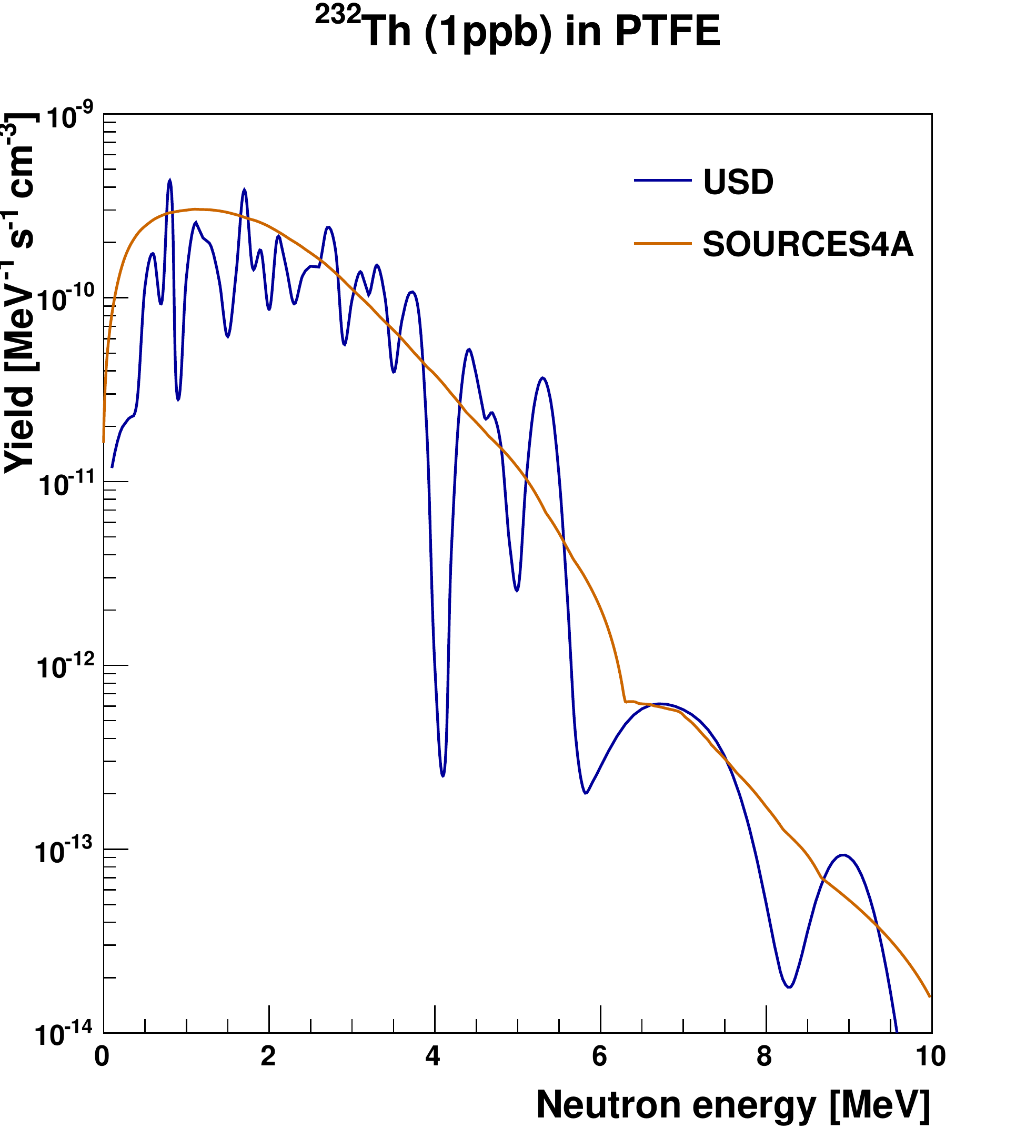} \\
\end{tabular}
\caption{Radiogenic neutron spectra (n$\cdot$MeV$^{-1}$$\cdot$s$^{-1}$$\cdot$cm$^{-3}$) calculated for 1~ppb $^{238}$U and $^{232}$Th decay chains, left and right panels, respectively. The ($\alpha$, n) reaction contribution is shown in orange for the SOURCES-4A  code and in blue for the USD webtool code. From top to bottom materials are polyethylene and teflon (PTFE).
\label{fig:allradspectra3}}
\end{figure*}

A comparison of neutron yield and energy spectra obtained via the SOURCES-4A and USD webtool codes for the same materials as above, has been performed. Results are shown in Table~\ref{tab:radneutron} and in Fig.~\ref{fig:allradspectra1}, Fig.~\ref{fig:allradspectra2} and Fig.~\ref{fig:allradspectra3}. A qualitative agreement between the two codes is observed. A maximum discrepancy of a factor 2 is found. The energy spectra calculated via SOURCES-4A code are in general smoother, without the presence of resonant peaks, a prominent feature of the USD spectra. 

To better estimate the discrepancies in the radiogenic neutron spectra due to the contribution of the cross section input library we have calculated radiogenic neutron yield and spectra for two materials: copper and polyethylene.  We have considered natural copper (70\% $^{63}$Cu and 30\% $^{65}$Cu) with a density of 8.96~g/cm$^{3}$; polyethylene material (C$_{2}$H$_{4}$) was considered with a density of density is 0.935~g/cm$^{3}$. Estimates are made assuming a concentration of 1~ppb of $^{238}$U and $^{232}$Th in their respective decay chains.  Cross sections in Cu are in good agreement in both codes, whereas the cross section of $^{13}$C  (in polythylene) shows some discrepancies, as illustrated in Fig.~\ref{fig:crosssection}. 

\begin{figure}[!htb]

\begin{tabular}{ll}\\
\includegraphics[width=5.5cm]{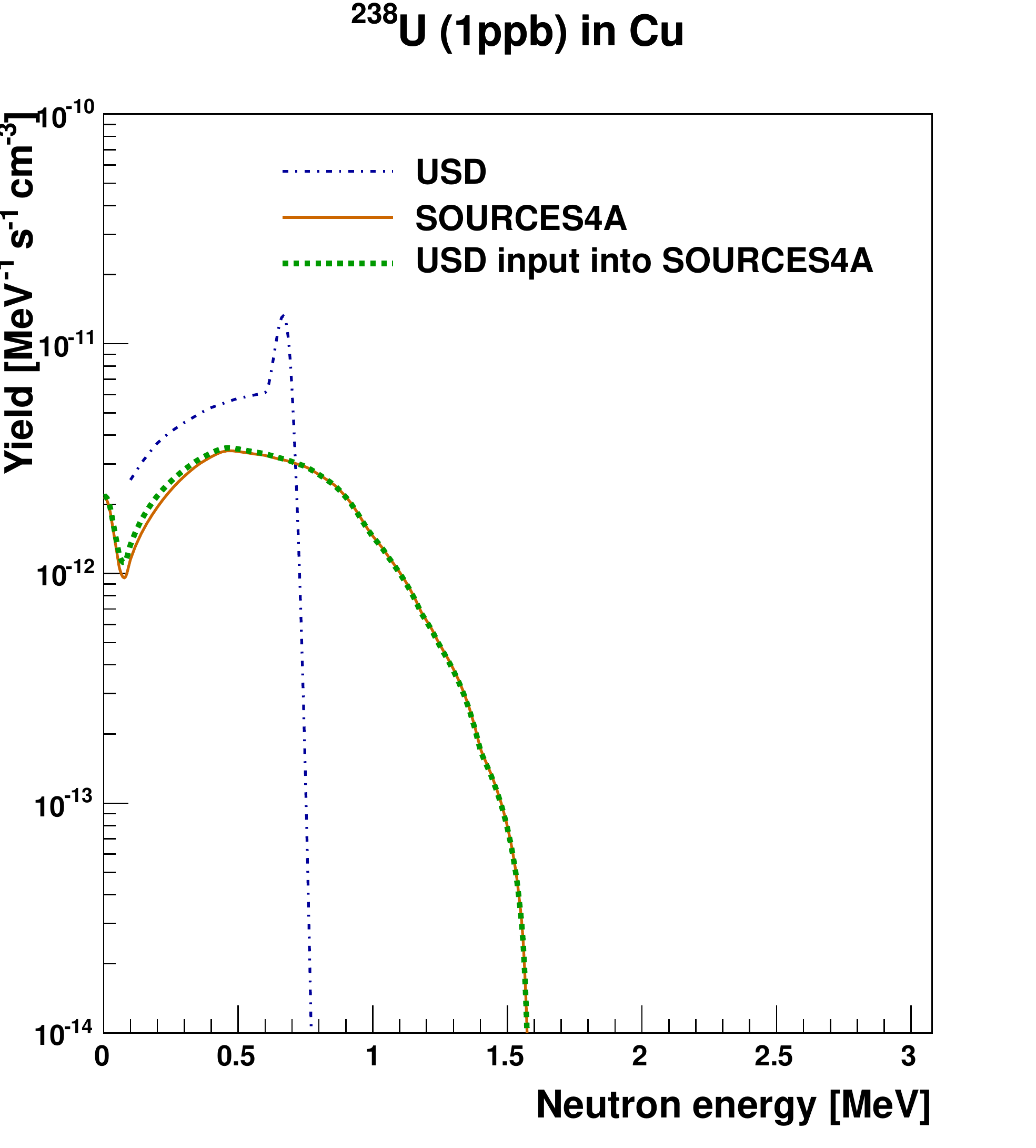} & \includegraphics[width=5.5cm]{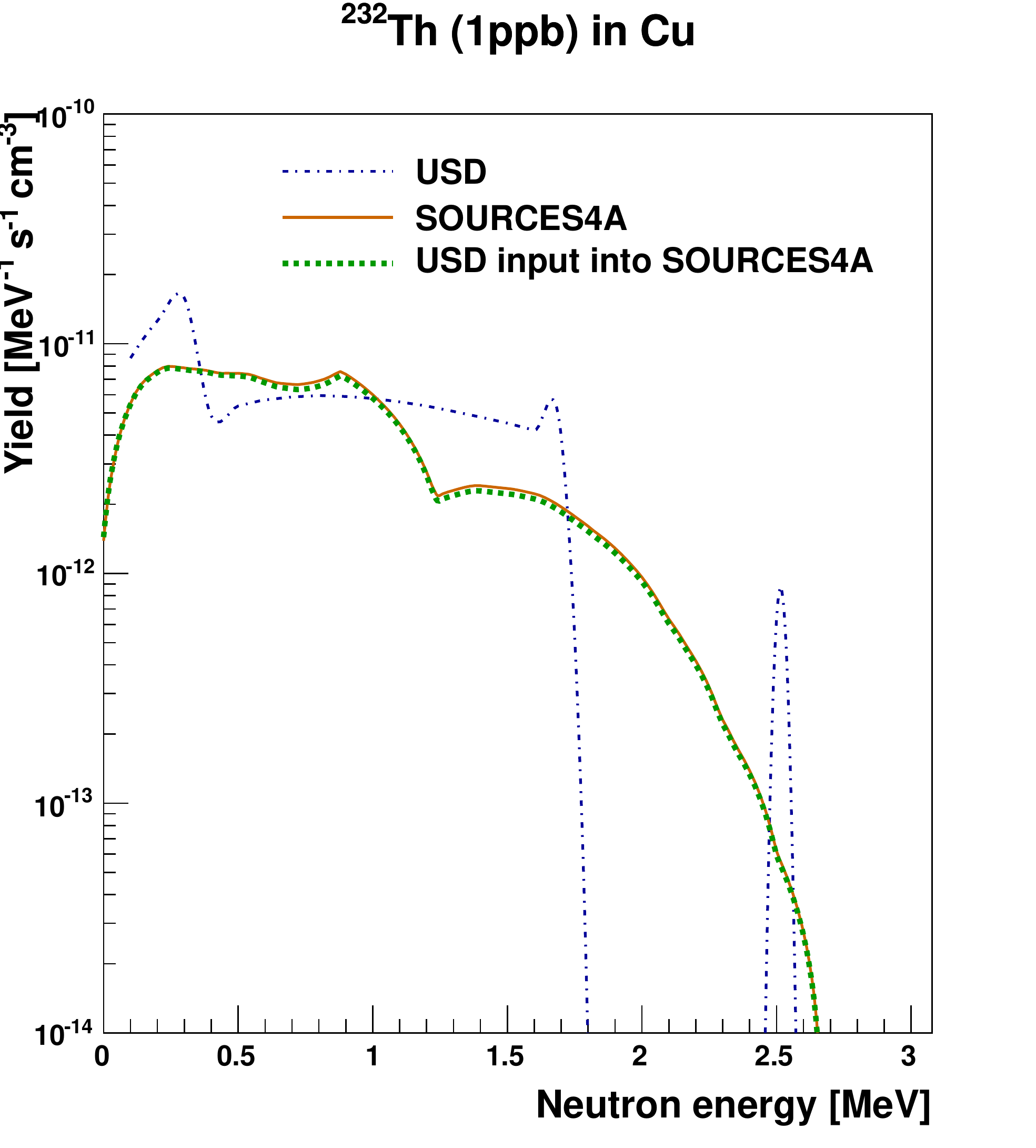} \\
\includegraphics[width=5.5cm]{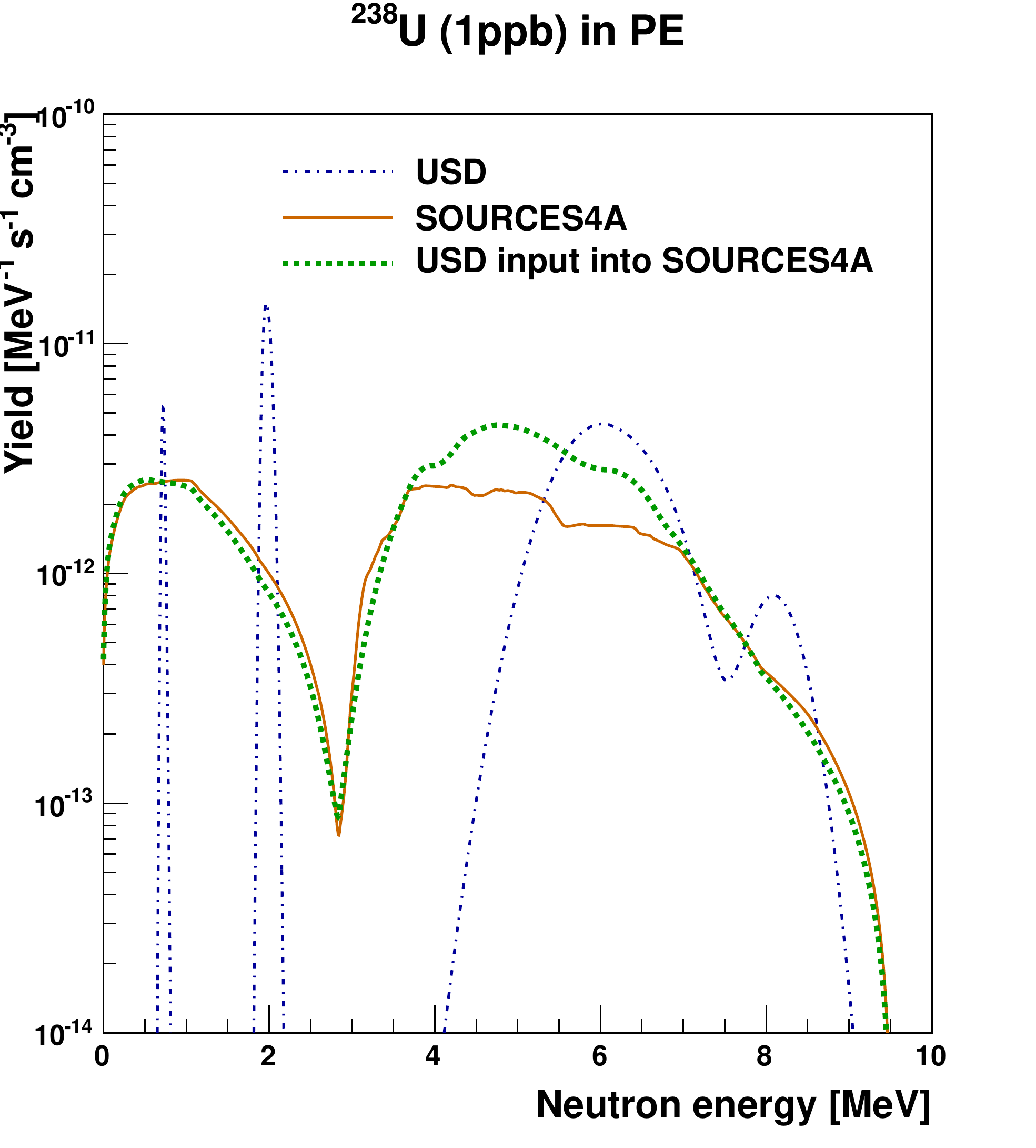} & \includegraphics[width=5.5cm]{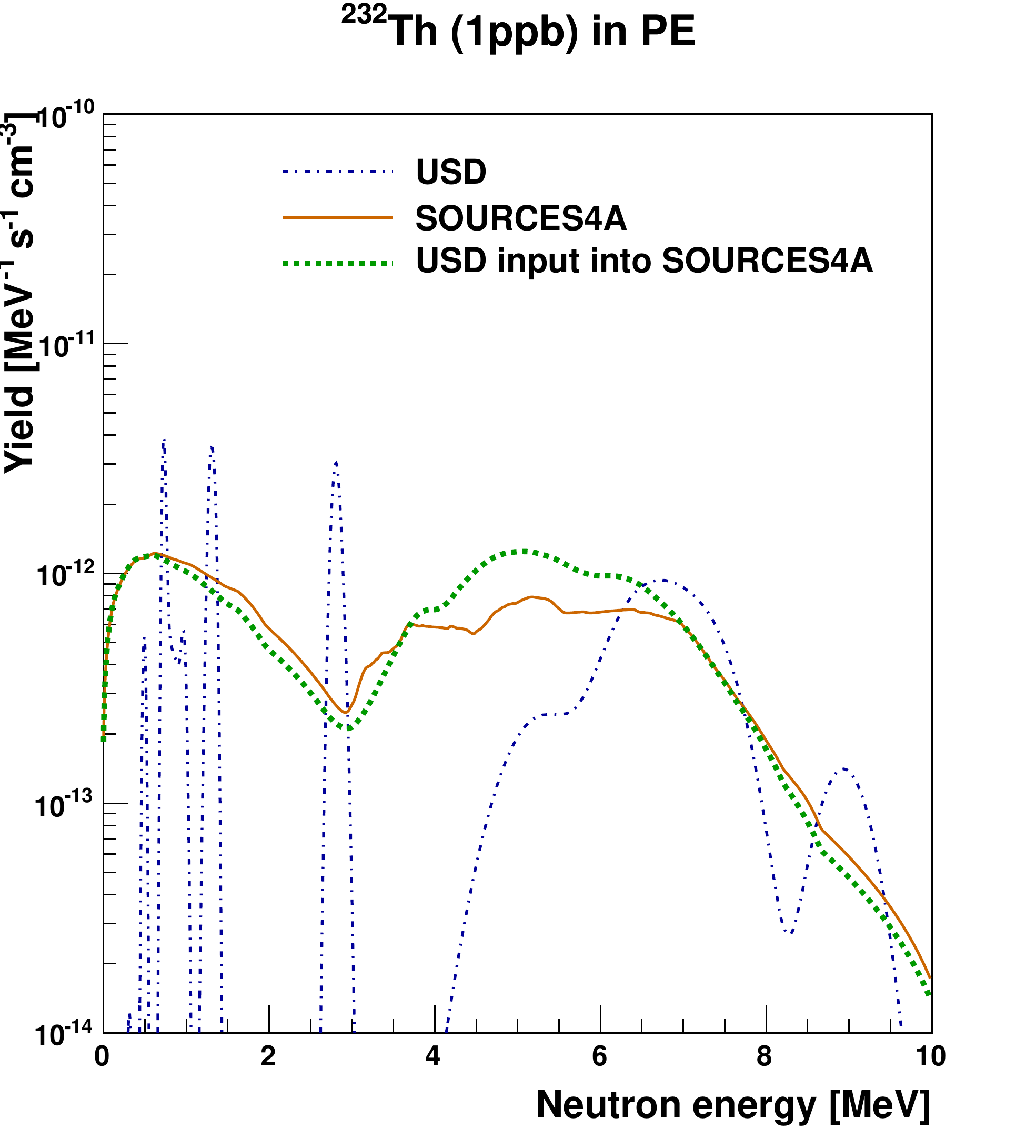} \\
\end{tabular}
\caption{Radiogenic neutron spectra (n$\cdot$MeV$^{-1}$$\cdot$s$^{-1}$$\cdot$cm$^{-3}$) calculated for 1~ppb $^{238}$U and $^{232}$Th decay chains, left and right panels, respectively. First row show copper contribution, lower row polyethylene material. Dotted blue lines refers to pure USD webtool code calculations, solid orange line to pure SOURCES-4A code calculations. Dashed green line reflects the computation of the SOURCES-4A algorithm with input cross section of the USD webtool code.
\label{fig:radspectra}}
\end{figure}

To check the overall effect that the input libraries had on the neutron yield results, the SOURCES-4A algorithm was used with the ($\alpha$,n) cross sections of the USD webtool code as input to calculate radiogenic neutron yields.  The results of this study are listed in  Table~\ref{tab:cu-poly}. 
The second to last column, (a) in Table~\ref{tab:cu-poly}, is the ratio of the radiogenic neutron yields obtained with the SOURCES-4A and USD webtool codes (column(2)/column(1)): the codes show a reasonably good agreement, within a 50\% discrepancy. The last column, (b) in Table~\ref{tab:cu-poly}, refers to the ratio of radiogenic neutron yield resulting from the USD webtool calculation over that resulting from the calculation using the SOURCES-4A algorithm with the USD webtool ($\alpha$, n) cross-sections as input.
For polyethylene, we can conclude that the input cross-section may account for up to a 20\% discrepancy in neutron yield. Figure~\ref{fig:radspectra} shows radiogenic neutron spectra for copper (upper row)  and polyethylene (lower row),  both from uranium and thorium decay chains, left and right panels respectively.


\section{GEANT4 Propagation}\label{sec:geant4prop}

To evaluate the impact of the varying neutron spectra produced by the SOURCES-4A and the USD webtool codes, simplified detector geometries were created within a RAT \cite{RAT} framework with GEANT4.9.5.p01 and the pertinent high precision neutron physics list utilizing cross sections from G4NDL3.14 for neutrons under 20 MeV. Four simulations, for neutrons from uranium and thorium, with the spectra from the SOURCES-4A and the USD webtool codes (shown in Fig.~\ref{fig:radspectra}) were run for each relevant material under study.

\begin{figure*}[!htb]
\begin{center}
\begin{tabular}{c}\\
 \includegraphics[width=6cm]{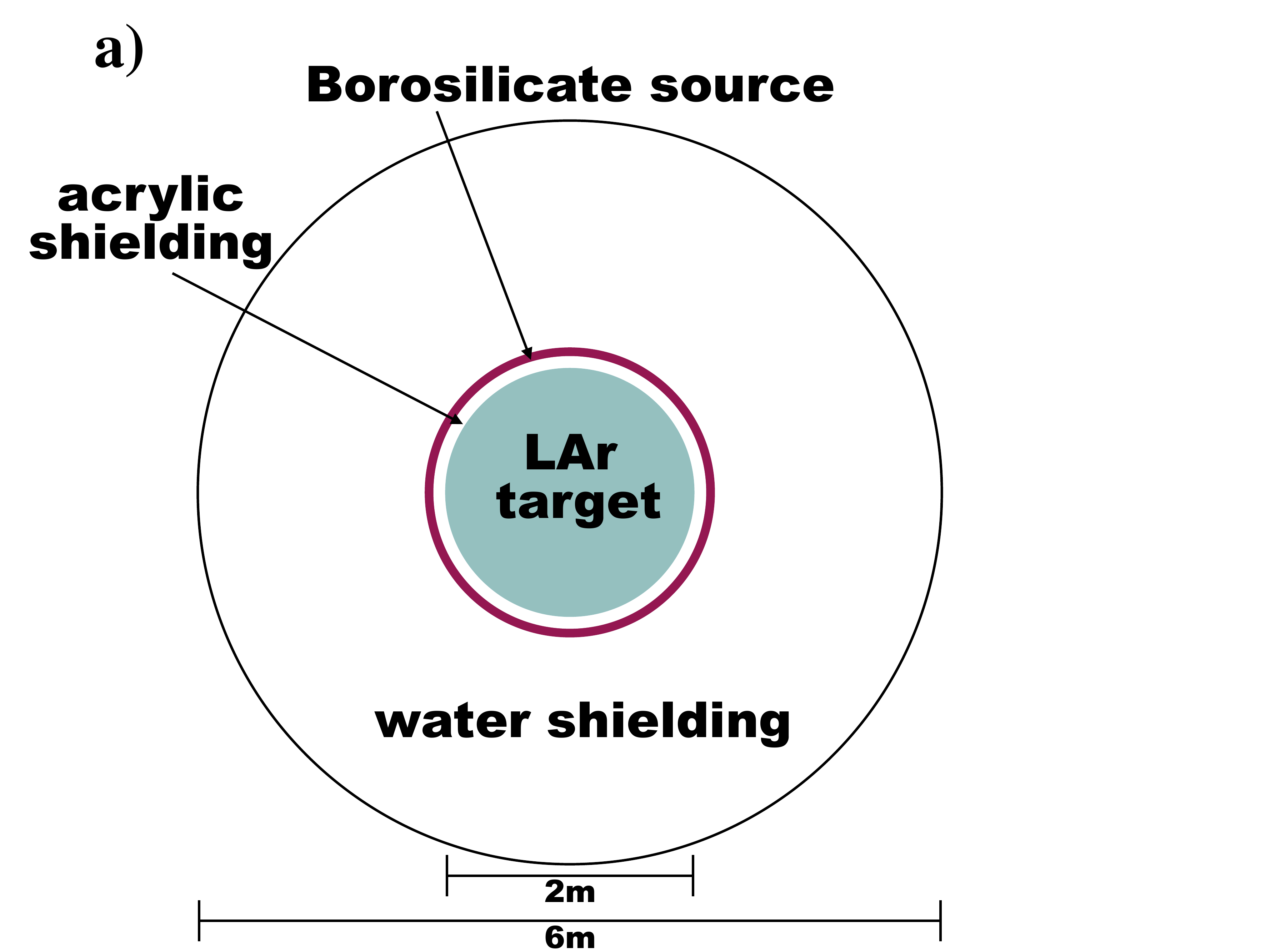}  \\ 
  \includegraphics[width=6cm]{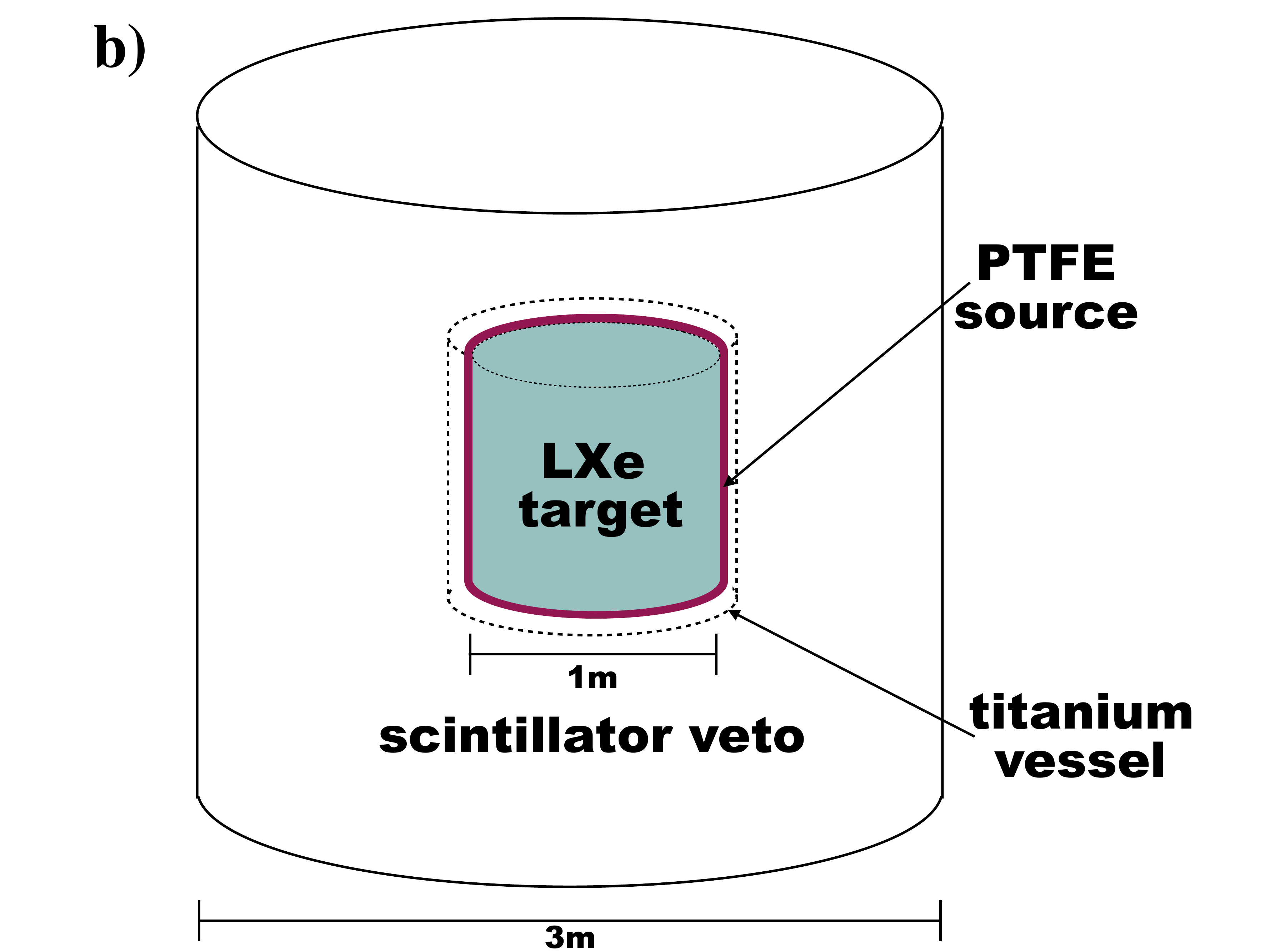} \\
  \includegraphics[width=6cm]{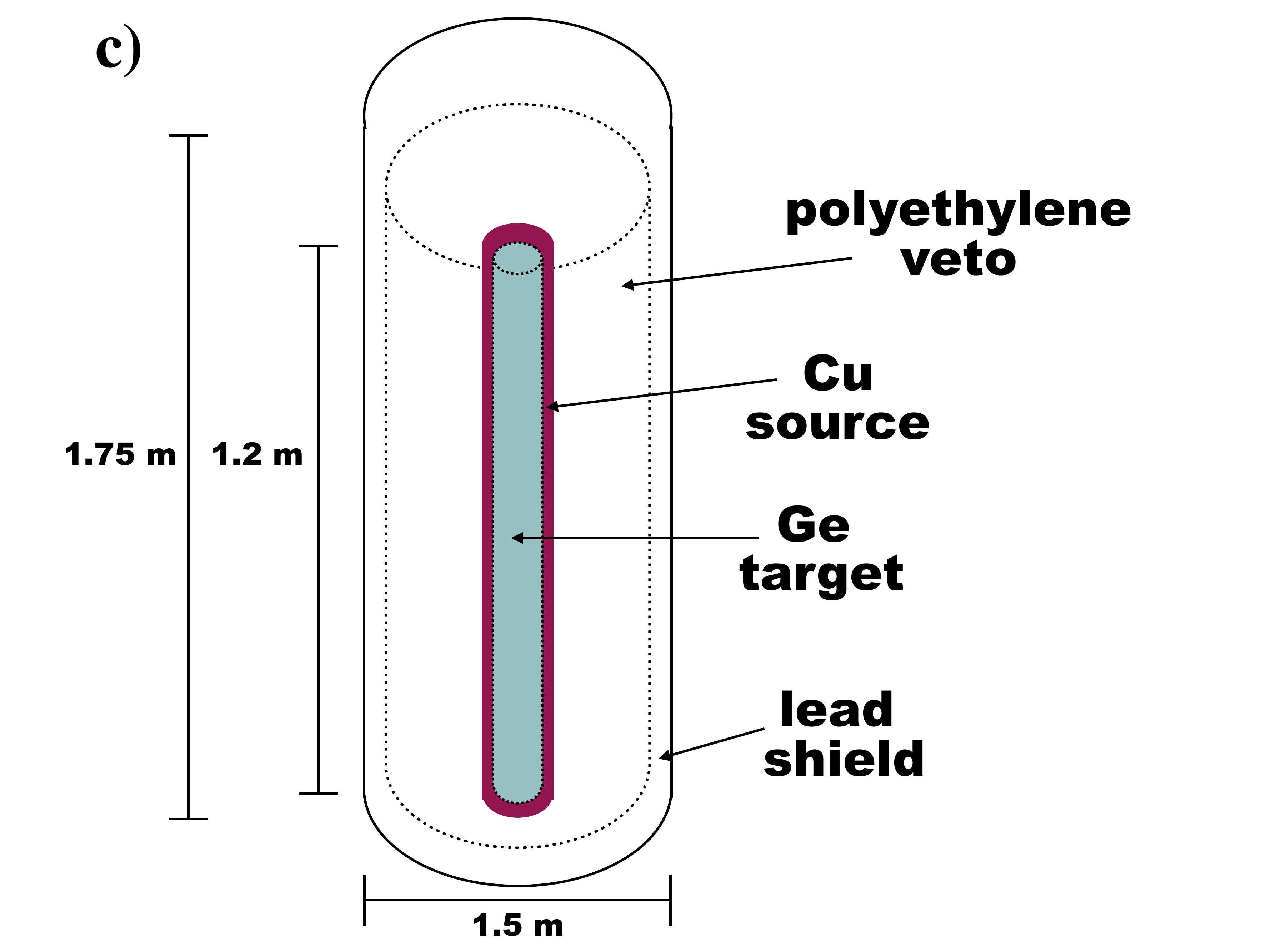} \\
\end{tabular}
\end{center}
\caption{The simplified geometries used in the neutron propagation studies, indicating the source material for the alpha-n neutrons in red, the detector target in green, and any shielding or vetoes as white. a) is the nested spherical geometry for a liquid argon target, b) is the nested square cylindrical geometry for a liquid xenon target with the cylinder diameters equal to their heights, and c) is the nested elongated cylinders for a germanium target. Additional shielding and dimensional details are in the text.
\label{fig:geodiagram}}
\end{figure*}

Three simplified direct dark matter detector geometries were established to study neutron elastic scatter signals within central detector materials and the potential for vetoing neutrons with outer detectors, depicted in Figure~\ref{fig:geodiagram}. The first was a spherical liquid argon detector, with a radius of 1$~$m, which is surrounded by shells of 10 cm thick acrylic, 5 mm thick borosilicate glass, and a water veto out to a radius of 3 $~$m. The simulated neutrons were isotropically created in the borosilicate glass, as it is the leading source of radiogenic neutrons in many liquid argon detectors.

The second geometry studied was a cylindrical liquid xenon detector with a 1$~$m diameter and height. It is nested within cylinders of 3$~$cm thick PTFE, 2$~$cm thick titanium, and liquid scintillator veto with a diameter and height of 3$~$m. Neutrons were generated isotropically in the PTFE; it is not likely to be the leading source of neutron backgrounds for most xenon detectors but may contribute significantly to the total radiogenic neutron yield.

The final geometry studied was a cylindrical solid germanium detector with a 10$~$cm diameter and 120$~$cm height. It is surrounded by nested cylinders: first 1$~$cm thick copper, then 15$~$cm of polyethylene veto and 10$~$cm of lead. The neutrons are generated isotropically within the copper.

\begin{figure}[!htb]
\begin{center}
\begin{tabular}{ll}\\

 \includegraphics[width=5.5cm]{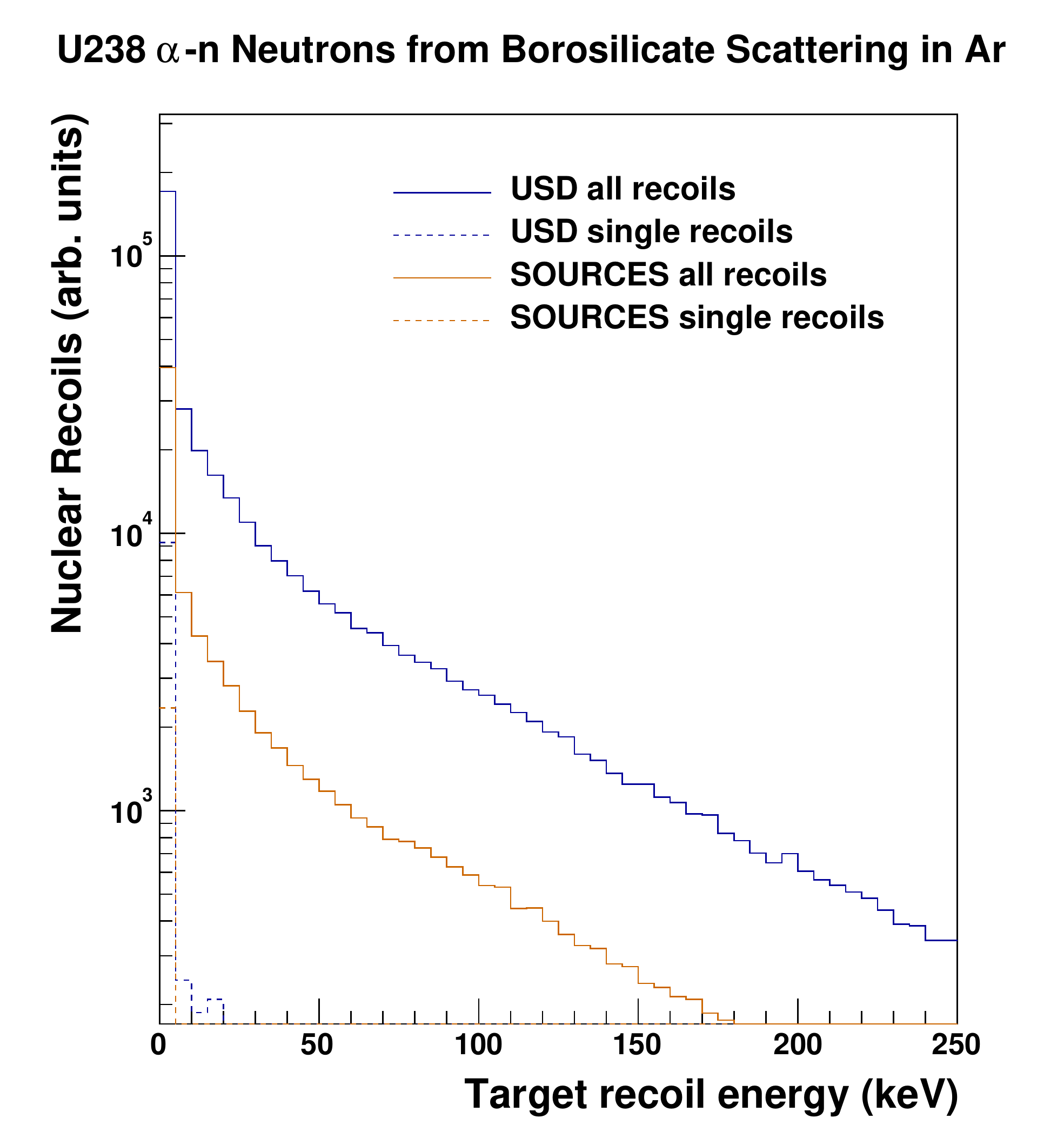} & \includegraphics[width=5.5cm]{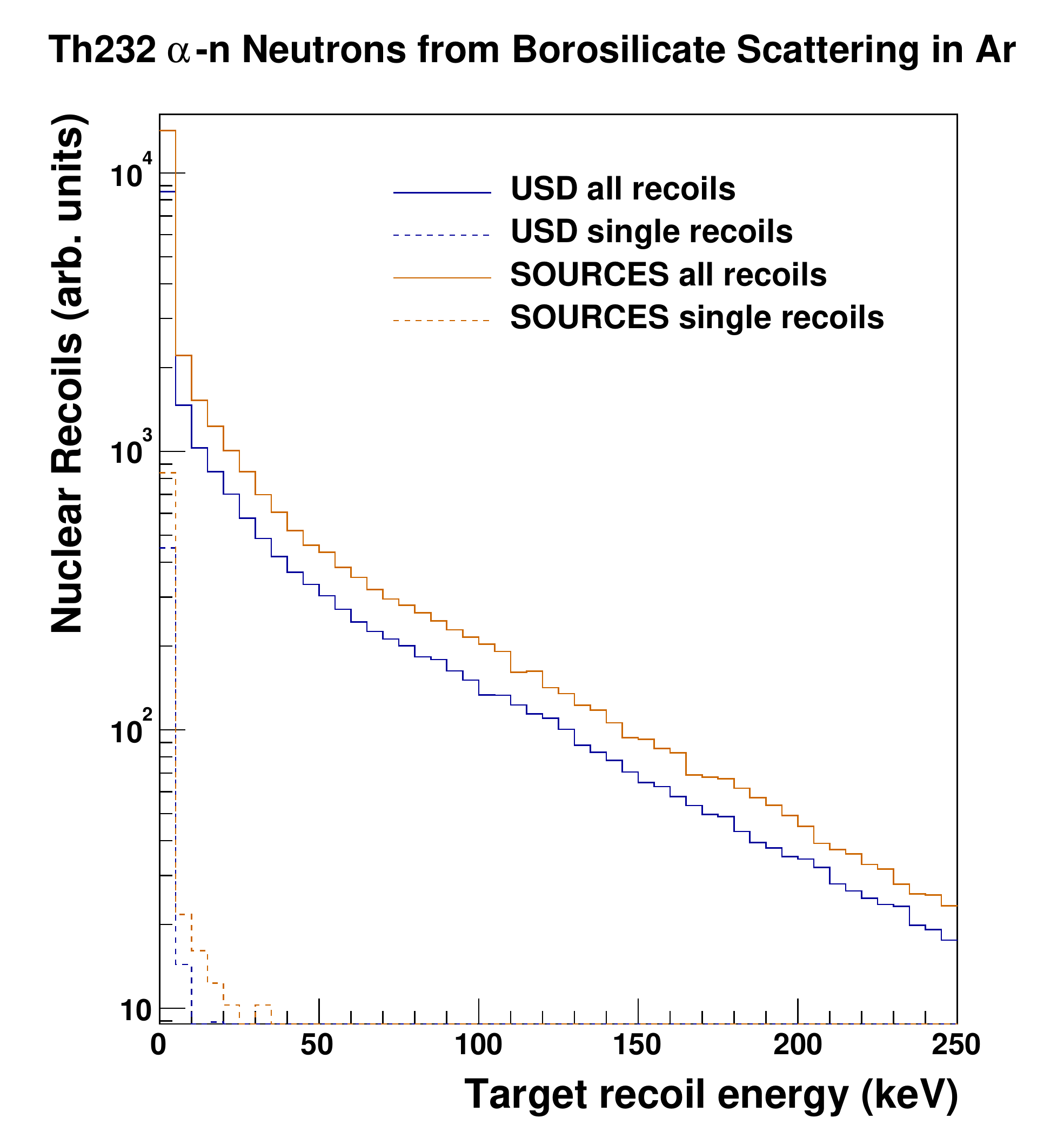} \\
 \includegraphics[width=5.5cm]{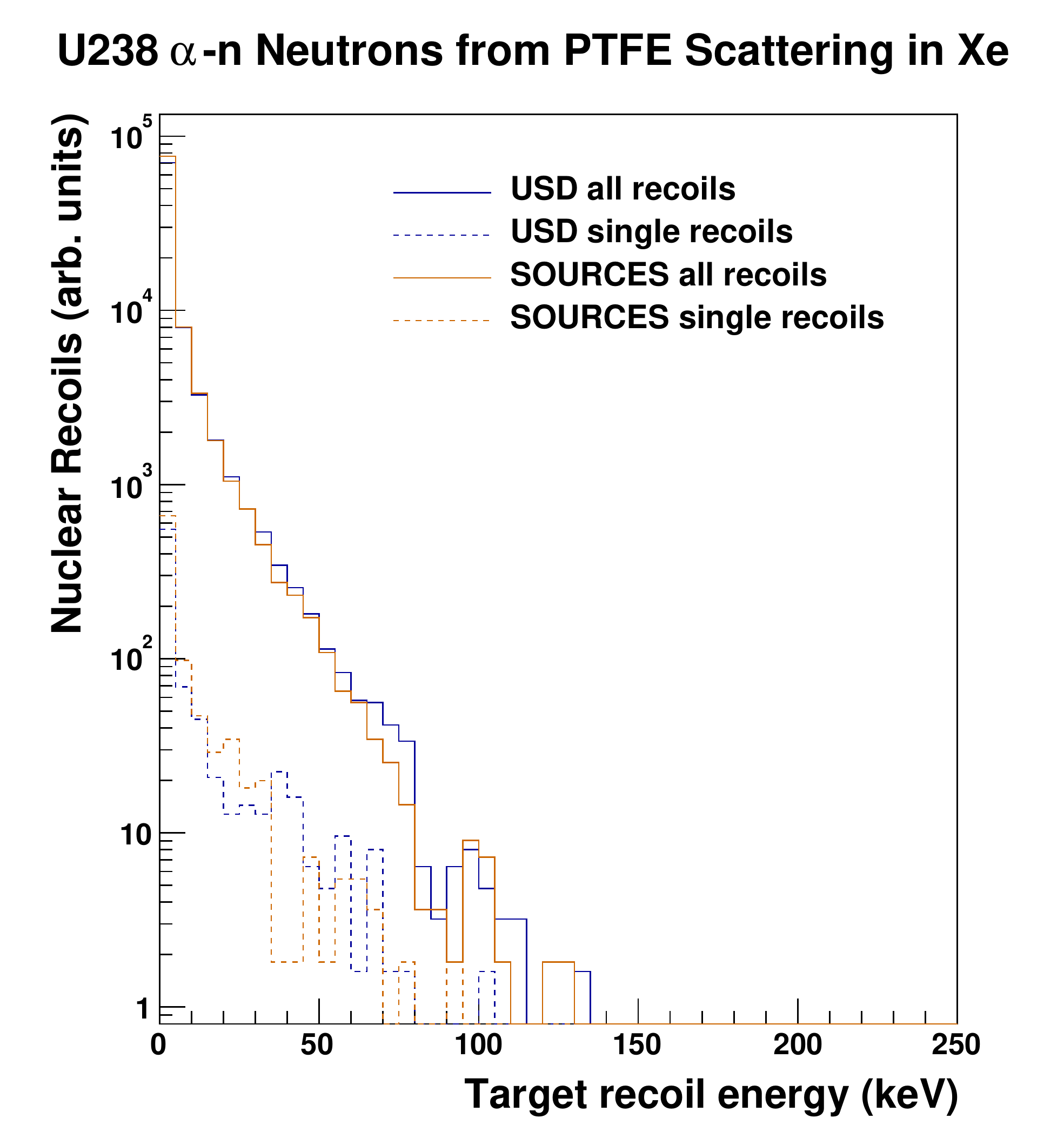} & \includegraphics[width=5.5cm]{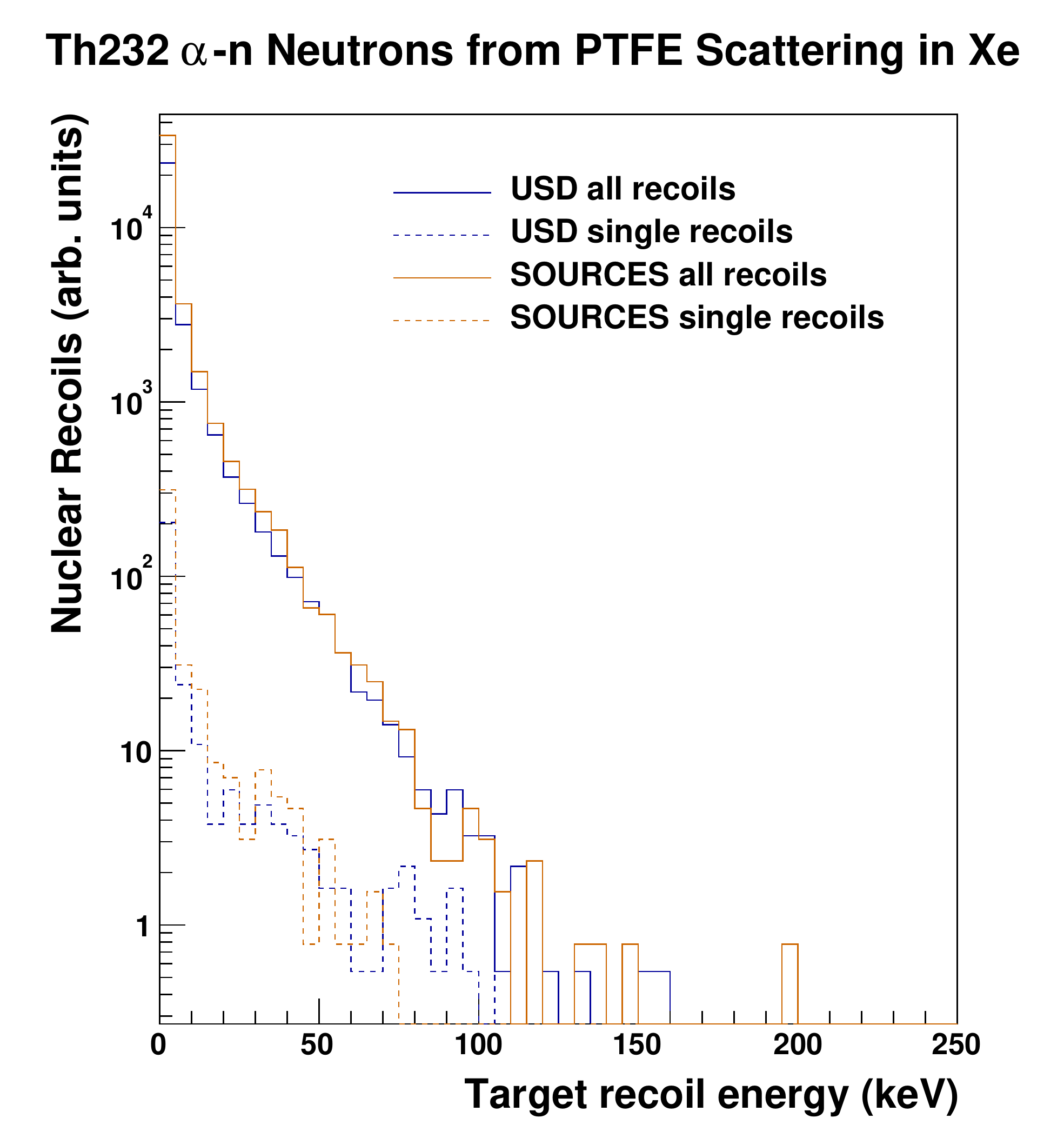} \\
 \includegraphics[width=5.5cm]{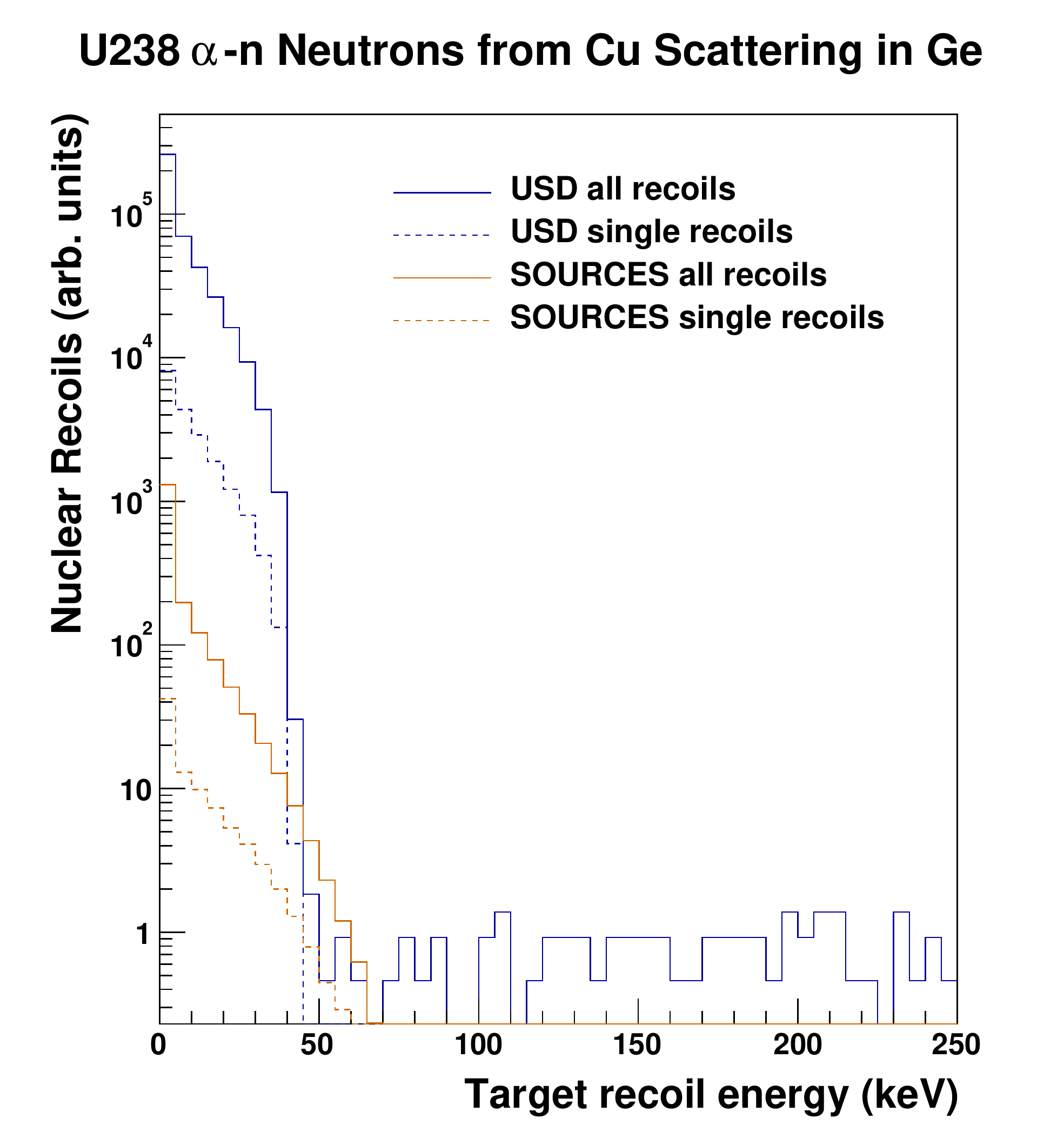} & \includegraphics[width=5.5cm]{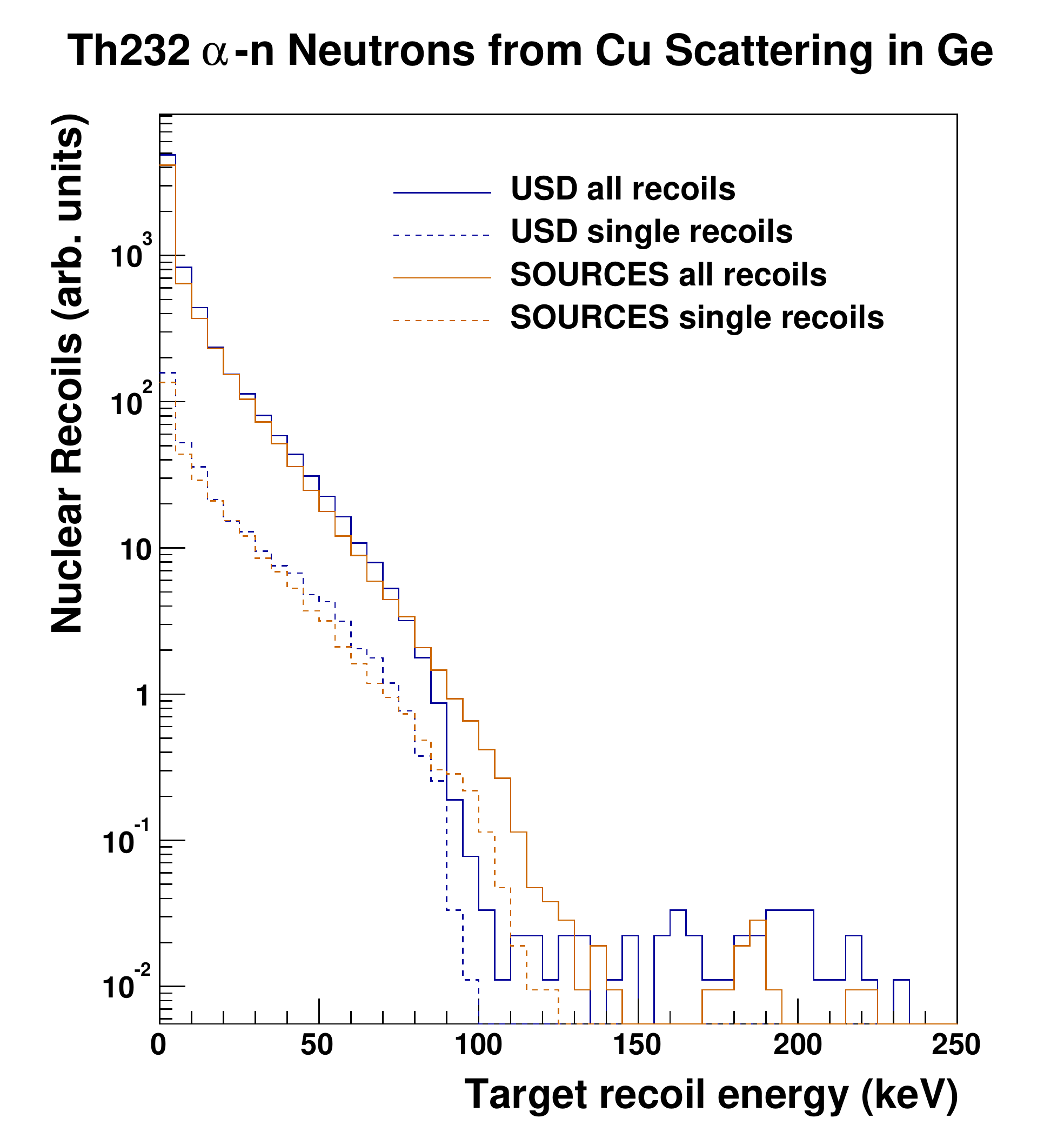} \\

\end{tabular}
\end{center}
\caption{Comparisons of nuclear recoils in simplified direct dark matter detector GEANT4 simulations induced from  ($\alpha$, n) neutrons originating in detector materials. All orange lines correspond to SOURCES4A initial spectra, while the USD initial spectra are plotted in blue. The solid lines are histograms of all individual nuclear recoils in the target materials, while the dashed lines are irreducible single nuclear recoils within the target. On the left are simulations for $^{238}$U, and $^{232}$Th spectra are on the right. The detectors from top to bottom are an argon target with neutrons originating in borosilicate glass, a xenon target with neutrons originating in PTFE, and a germanium target with neutrons originating in copper.
\label{fig:nuclearrecoils}}
\end{figure}

For these sample studies, an analysis threshold on the neutron-induced nuclear recoils of 20$~$keV was set in the argon detector and 5$~$keV in the xenon and germanium  detectors. Scatters were rejected as WIMP-like recoils if there were multiple distinguishable nuclear scatters over threshold within the target. Figure~\ref{fig:nuclearrecoils} shows the total induced nuclear recoil spectra in these simulated detector targets along with the single nuclear recoil spectra. The larger liquid noble detectors show greater reductions from all nuclear scatters to single nuclear scatters due to their size. The induced recoil spectra visibly smooth out the shape differences between the input neutron yield spectra, and the differences between simulations originating with the SOURCES-4A and USD webtool codes are minimized when studying the single nuclear recoils of interest.

Table~\ref{tab:recoilcounts} provides the percentage difference between the simulated nuclear recoil counts from the  neutron spectra. The difference in counts for the borosilicate and PTFE studies can nearly all be attributed to the difference in the total neutron yields. Indeed, these count differences are generally smoothed out and reduced proportionately from the yield differences in Table~\ref{tab:radneutron}, and a $\chi^2$ per degree of freedom test of just the recoil spectra shapes shows reasonable agreement between both simulations.

This is not the case for the neutrons originating in copper. The total yields began with a 20\% agreement, but the truncation in energy of the USD spectra causes a significant difference of up to 80\% for the numbers of recoils seen above threshold. These differences are easily seen in the lower-left panel of Fig.~\ref{fig:nuclearrecoils} for the $^{238}$U neutrons originating in Cu and recoiling in a Ge target.

Additional tests, that are not shown here, vetoed events with more than 1$~$keV deposited from inelastic or capture gamma ray scatters within the target, or if a neutron capture occurred within the veto material. Although no common dopants were included in the veto materials, most neutrons did capture within the vetoes. The remaining elastic nuclear scatters comparisons between the SOURCES-4A and USD webtool codes initial spectra were analogous to those for the single recoils, which is why we have chosen not to add them to the plots of Fig.~\ref{fig:nuclearrecoils}. 

The impact upon neutron background simulations for dark matter detectors of the difference between ($\alpha$, n) neutron spectra calculated with the SOURCES-4A and the USD webtool is primarily one of overall normalization. The differences would lead to different background predictions prior to running an experiment, but when spectral fits are made to recoils seen in a detector for multiple scatters, high energy or high radius events, the prediction of low energy single nuclear recoils is quite robust. However, there may be other exceptions besides copper to these spectral considerations.

\section{Conclusion}\label{sec:concl}

The low radioactive background physics community has access to two tools for calculating $\alpha$-n neutron yield spectra for common $\alpha$-decay sources: SOURCES-4A and the USD webtool. Both codebases may meet the needs of particular users. 

SOURCES-4A allows the user full control of the reactions they are studying, including the ability to study the decay chains out of secular equilibrium between their early and late part of the chain. In addition, the neutron yields from spontaneous fission are also easily calculated. However, personal correspondence with the code developers s advised for updates on extended energy ranges and corrected cross sections.

The USD webtool provides a user-friendly webform interface to quickly obtain neutron spectra with realistic resonant peaks. The lack of fission yields and options for broken equilibrium are disadvantages compared to the customizable SOURCES-4A.

Between the two tools we found no systematic differences between the input cross-sections and  output spectra and yields. Both may have errors in cross sections or outputs that require a human eye to catch. Low energy neutron physics codes been notoriously difficult to benchmark, and the agreement to 50\% or better between these packages can probably be interpreted as bracketing the known range of neutron yields.

Once the neutron yields are used as the input to background simulations, the differences in both yield and spectral shape are smoothed away in GEANT4 Monte Carlo studies of neutron induced nuclear recoils. As common additional cuts are placed on single nuclear recoils without additional gamma ray signals from inelastic scatters or neutron capture in a veto, both SOURCES-4A and USD input spectra predict similar background counts.

A complete comparison and validation of these tools would require comparison with data. However, such data sets are difficult to obtain. For a running experiment, detailed geometries of fully assayed parts would be necessary to compare against recoil spectra generated by SOURCES-4A and USD. Statistical agreement with both code bases is likely with the purposefully low rates of neutron recoils within low background experiments.

SOURCES-4A has a long history of use within the low background community, and will continue to be used for simulating future generations of experiments. As the newer TALYS nuclear code base that USD relies upon is exercised in other nuclear physics settings, the greater the likelihood of the USD webtool or a similar TALYS based calculation being used for radiogenic background predictions.  Both will offer benefits to their users.  In addition new tools and codes such as  ~\cite{Westerdale:2017kml} will continue to be developed as new measurements become available.

\newpage
\section{Acknowledgement}

This work was supported by the National Science Foundation under grant No. PHY 124260 awarded to the AARM (Assay and Acquisition of Radiopure Materials) collaboration.  Any opinions, findings, and conclusions or recommendations expressed in this material are those of the author(s) and do not necessarily reflect the views of the National Science Foundation.  This material is further based upon work supported by the U.S. Department of Energy, Office of Science, under Award Number DE-SC0015652. We acknowledge Vitaly Kudryavtsev for providing us updated improvements to the SOURCES code and libraries and Dongming Mei for providing online access to the USD neutron yield code.  Crucial computing support provided by the SMU High Performance Computing Cluster is acknowledged gratefully.

\nocite{*}

\section*{References}
\bibliographystyle{elsarticle-num}
\bibliography{radiogenicNotes}

\begin{thebibliography}{10}
\expandafter\ifx\csname url\endcsname\relax
  \def\url#1{\texttt{#1}}\fi
\expandafter\ifx\csname urlprefix\endcsname\relax\def\urlprefix{URL }\fi
\expandafter\ifx\csname href\endcsname\relax
  \def\href#1#2{#2} \def\path#1{#1}\fi

\bibitem{Jungman:1995df}
G.~Jungman, M.~Kamionkowski, K.~Griest, {Supersymmetric dark matter}, Phys.
  Rept. 267 (1996) 195--373.
\newblock \href {http://arxiv.org/abs/hep-ph/9506380}
  {\path{arXiv:hep-ph/9506380}}, \href
  {http://dx.doi.org/10.1016/0370-1573(95)00058-5}
  {\path{doi:10.1016/0370-1573(95)00058-5}}.

\bibitem{Heaton:1989vbt}
R.~Heaton, H.~Lee, P.~Skensved, B.~C. Robertson, {Neutron production from
  thick-target (alpha, n) reactions}, Nucl. Instrum. Meth. A276~(3) (1989)
  529--538.
\newblock \href {http://dx.doi.org/10.1016/0168-9002(89)90579-2}
  {\path{doi:10.1016/0168-9002(89)90579-2}}.

\bibitem{Formaggio:2004ge}
J.~A. Formaggio, C.~J. Martoff, {Backgrounds to sensitive experiments
  underground}, Ann. Rev. Nucl. Part. Sci. 54 (2004) 361--412.
\newblock \href {http://dx.doi.org/10.1146/annurev.nucl.54.070103.181248}
  {\path{doi:10.1146/annurev.nucl.54.070103.181248}}.

\bibitem{Akerib:2014rda}
D.~S. Akerib, et~al., {Radiogenic and Muon-Induced Backgrounds in the LUX Dark
  Matter Detector}, Astropart. Phys. 62 (2015) 33--46.
\newblock \href {http://arxiv.org/abs/1403.1299} {\path{arXiv:1403.1299}},
  \href {http://dx.doi.org/10.1016/j.astropartphys.2014.07.009}
  {\path{doi:10.1016/j.astropartphys.2014.07.009}}.

\bibitem{Cai:2011zzc}
B.~Cai, M.~Boulay, B.~Cleveland, T.~Pollmann, {Surface backgrounds in the
  DEAP-3600 dark matter experiment}, AIP Conf. Proc. 1338 (2011) 137--146.
\newblock \href {http://dx.doi.org/10.1063/1.3579572}
  {\path{doi:10.1063/1.3579572}}.

\bibitem{Aprile:2015uzo}
E.~Aprile, et~al., {Physics reach of the XENON1T dark matter experiment}, JCAP
  1604~(04) (2016) 027.
\newblock \href {http://arxiv.org/abs/1512.07501} {\path{arXiv:1512.07501}},
  \href {http://dx.doi.org/10.1088/1475-7516/2016/04/027}
  {\path{doi:10.1088/1475-7516/2016/04/027}}.

\bibitem{Akerib:2015cja}
D.~S. Akerib, et~al., {LUX-ZEPLIN (LZ) Conceptual Design Report }\href
  {http://arxiv.org/abs/1509.02910} {\path{arXiv:1509.02910}}.

\bibitem{Agnese:2016cpb}
R.~Agnese, et~al., {Projected Sensitivity of the SuperCDMS SNOLAB experiment},
  Phys. Rev. D95~(8) (2017) 082002.
\newblock \href {http://arxiv.org/abs/1610.00006} {\path{arXiv:1610.00006}},
  \href {http://dx.doi.org/10.1103/PhysRevD.95.082002}
  {\path{doi:10.1103/PhysRevD.95.082002}}.

\bibitem{Wright:2010bu}
A.~Wright, P.~Mosteiro, B.~Loer, F.~Calaprice, {A Highly Efficient Neutron Veto
  for Dark Matter Experiments}, Nucl. Instrum. Meth. A644 (2011) 18--26.
\newblock \href {http://arxiv.org/abs/1010.3609} {\path{arXiv:1010.3609}},
  \href {http://dx.doi.org/10.1016/j.nima.2011.04.009}
  {\path{doi:10.1016/j.nima.2011.04.009}}.

\bibitem{Calkins:2015hya}
R.~Calkins, B.~Loer, {Prototyping an Active Neutron Veto for SuperCDMS}, AIP
  Conf. Proc. 1672 (2015) 140002.
\newblock \href {http://arxiv.org/abs/1506.01922} {\path{arXiv:1506.01922}},
  \href {http://dx.doi.org/10.1063/1.4928018} {\path{doi:10.1063/1.4928018}}.

\bibitem{Loach:2016fsk}
J.~C. Loach, J.~Cooley, G.~A. Cox, Z.~Li, K.~D. Nguyen, A.~W.~P. Poon, {A
  Database for Storing the Results of Material Radiopurity Measurements}, Nucl.
  Instrum. Meth. A839 (2016) 6--11.
\newblock \href {http://arxiv.org/abs/1604.06169} {\path{arXiv:1604.06169}},
  \href {http://dx.doi.org/10.1016/j.nima.2016.09.036}
  {\path{doi:10.1016/j.nima.2016.09.036}}.

\bibitem{Wilson2009608}
W.~Wilson, R.~Perry, W.~Charlton, T.~Parish,
  \href{http://www.sciencedirect.com/science/article/pii/S0149197008001418}{Sources:
  A code for calculating (alpha, n), spontaneous fission, and delayed neutron
  sources and spectra}, Progress in Nuclear Energy 51~(4–5) (2009) 608 --
  613.
\newblock \href
  {http://dx.doi.org/http://dx.doi.org/10.1016/j.pnucene.2008.11.007}
  {\path{doi:http://dx.doi.org/10.1016/j.pnucene.2008.11.007}}.
\newline\urlprefix\url{http://www.sciencedirect.com/science/article/pii/S0149197008001418}

\bibitem{Mei:2008ir}
D.~M. Mei, C.~Zhang, A.~Hime, {Evaluation of (alpha,n) Induced Neutrons as a
  Background for Dark Matter Experiments}, Nucl. Instrum. Meth. A606 (2009)
  651--660.
\newblock \href {http://arxiv.org/abs/0812.4307} {\path{arXiv:0812.4307}},
  \href {http://dx.doi.org/10.1016/j.nima.2009.04.032}
  {\path{doi:10.1016/j.nima.2009.04.032}}.

\bibitem{Selvi:2013nru}
M.~Selvi, {Review of Monte Carlo simulations for backgrounds from
  radioactivity}, AIP Conf. Proc. 1549 (2013) 213--218.
\newblock \href {http://dx.doi.org/10.1063/1.4818111}
  {\path{doi:10.1063/1.4818111}}.

\bibitem{VK}
{Private conversation with Vitaly Kudryavtsev (University of Sheffield, UK)},
  v.kudryavtsev@sheffield.ac.uk.

\bibitem{EMPIRE}
{EMPIRE-II: statistical model code for nuclear reaction calculation},
  https://www-nds.iaea.org/empire218/manual.ps.

\bibitem{Tomasello:2006sm}
V.~Tomasello, V.~A. Kudryavtsev, {Calculation of neutron yield from
  radioactivity in materials relevant to dark matter searches}, in:
  {Proceedings, 6th International Workshop on The identification of dark matter
  (IDM 2006): Rhodes, Greece, September 11-16, 2006}, 2006, pp. 537--541.

\bibitem{Jacobs:1983}
G.~J.~H. Jacobs, H.~Liskien, {Energy Spectra of Neutrons Produced by a-
  Particles in Thick Targets of Light Elements}, Ann. Nucl. Energy 10 (1983)
  541.

\bibitem{TENDL12}
A.~Koning, D.~Rochman, S.~van~der Marck, J.~Kopecky, J.~C. Sublet, S.~Pomp,
  H.~Sjostrand, R.~Forrest, E.~Bauge, H.~Henriksson, {}, TENDL-2012:
  TALYS-based evaluated nuclear data library,
  ftp://ftp.nrg.eu/pub/www/talys/tendl2012/tendl2012.html.

\bibitem{TALYS}
A.~Koning, D.~Rochman, {TENDL-2012: TALYS-based evaluated nuclear data
  library}, Nuclear Research and Consultancy Group (NRG),
  http://www.talys.eu/documentation/.

\bibitem{RAT}
Rat (is an analysis tool) user s guide,
  \url{http://rat.readthedocs.org/en/latest/}, accessed: 2016-01-21.

\bibitem{Westerdale:2017kml}
S.~Westerdale, P.~D. Meyers, {Radiogenic Neutron Yield Calculations for
  Low-Background Experiments}\href {http://arxiv.org/abs/1702.02465}
  {\path{arXiv:1702.02465}}.

\end{thebibliography}

\section*{}

\newpage

\newpage
\newpage
\setlength\tabcolsep{15pt}
\begin{table}
\caption{Alpha lines present in the SOURCES-4A and USD webtool calculations, and their intensity (BR) for isotopes in the $^{232}$Th and  $^{238}$U decay chains. Only lines with intensity $>1\%$ have been quoted.}
\label{tab:alphaline}
\begin{tabular}{*{5}{c}}
&  \multicolumn{2}{c}{\textbf{SOURCES-4A}} & \multicolumn{2}{c}{\textbf{USD}}\\
\textbf{Isotopes} & {\footnotesize \textbf{Line (keV)} }& {\footnotesize \textbf{BR (\%)}} & {\footnotesize \textbf{Line (keV)}} & {\footnotesize \textbf{BR (\%)}} \\ 
\hline
\hline
\multirow{2}{*}{\textbf{$^{238}$U}} & 4151 & 21 & 4151 & 21\\
& 4198 & 79 & 4198 & 79\\
\cline{2-5}
\multirow{2}{*}{\textbf{$^{234}$U}} & 4722.4 & 28.42 & 4722 & 28.6\\
& 4774.6 & 71.38 & 4775 & 71.4\\
\cline{2-5}
\multirow{2}{*}{\textbf{$^{230}$Th}} & 4620.5 & 23.4 & 4621 & 23.7\\
& 4687.0 & 76.3 & 4688 & 76.3\\
\cline{2-5}
\multirow{2}{*}{\textbf{$^{226}$Ra}} & 4601 & 5.55 & 4602 & 5.6\\
& 4784.34 & 94.45 & 4784 & 94.4\\
\cline{2-5}
\multirow{1}{*}{\textbf{$^{222}$Rn}} & 5489 &99.92 & 5490 & 100\\
\cline{2-5}
\multirow{1}{*}{\textbf{$^{218}$Po}} & 6002.35 & 99.98 & 6002 & 100\\
\cline{2-5}
\multirow{1}{*}{\textbf{$^{214}$Po}} & 7686.82 & 99.99 & 7687 & 99.99\\
\cline{2-5}
\multirow{1}{*}{\textbf{$^{210}$Po}} &5304.33 & 99.99 & 5304 & 100\\
\cline{1-5}
\multirow{2}{*}{\textbf{$^{232}$Th}} & 3947.2 & 21.7 & 3954 & 22.1\\
& 4012.3 & 78.2 & 4013 & 77.9\\
\cline{2-5}
\multirow{2}{*}{\textbf{$^{228}$Th}} & 5340.36 & 27.2 & 5340 & 28.5\\
& 5423.15 & 72.2 & 5.423 & 71.5\\
\cline{2-5}
\multirow{2}{*}{\textbf{$^{224}$Ra}} & 5448.6 & 5.06 & 5449 & 5.1\\
& 5685.37 & 94.92 & 5.685 & 94.9\\
\cline{2-5}
\multirow{1}{*}{\textbf{$^{220}$Rn}} & 6288.3 & 99.99 & 6288 & 100\\
\cline{2-5}
\multirow{1}{*}{\textbf{$^{216}$Po}} & 6778.5 & 99.99 & 6778 & 100\\
\cline{2-5}
\multirow{2}{*}{\textbf{$^{212}$Bi}} & 6051.1 & 25.16 & 6050 & 26.2\\
& 6090.2 & 9.79 & 6090.2 & 9.8\\
\cline{2-5}
\multirow{1}{*}{\textbf{$^{212}$Po}} & 8784.6 & 100 & 8784 & 64\\

\end{tabular}

\end{table}%

\setlength\tabcolsep{10pt}
\begin{table}
\caption{Radiogenic neutron yield (n$\cdot$s$^{-1}$$\cdot$cm$^{-3}$) from ($\alpha$, n) reactions in different materials considering the $^{238}$U and $^{232}$Th decay chains.  Neutron yields have been calculated for concentrations of 1~ppb  $^{238}$U and $^{232}$Th using the modified SOURCES-4A code. } 
\label{tab:contributions}
\begin{center}
\begin{tabular}{lcc}
&  \multicolumn{2}{c}{\textbf{Neutron yield for 1~ppb (n$\cdot$s$^{-1}$$\cdot$cm$^{-3}$)}} \\
\textbf{Material} &  \textbf{$^{238}$U $\rightarrow$ $^{226}$Ra} & \textbf{$^{226}$Ra $\rightarrow$ $^{206}$Pb} \\ 
\hline
\hline
Stainless Steel &  6.4~$\cdot$~10$^{-15}$ & 3.1~$\cdot$~10$^{-11}$\\
Pyrex &  4.0~$\cdot$~10$^{-11}$ & 1.9~$\cdot$~10$^{-10}$\\
Borosilicate Glass   & 6.3~$\cdot$~10$^{-11}$ & 2.8~$\cdot$~10$^{-10}$\\
Titanium  &  1.14~$\cdot$~10$^{-13}$ & 1.0~$\cdot$~10$^{-10}$\\
Copper    &  0.0~$\cdot$~10$^{-11}$ & 2.8~$\cdot$~10$^{-12}$\\
PE (C$_{2}$H$_{4}$) &  1.6~$\cdot$~10$^{-12}$ & 1.1~$\cdot$~10$^{-11}$\\ 
PTFE (CF$_{2}$) & 1.8~$\cdot$~10$^{-10}$ & 1.6~$\cdot$~10$^{-9}$\\
\hline
& \textbf{ $^{232}$Th $\rightarrow$ $^{228}$Th} & \textbf{$^{228}$Th $\rightarrow$ $^{208}$Pb} \\ 
\hline
\hline
Stainless Steel & 8.8~$\cdot$~10$^{-19}$ & 4.1~$\cdot$~10$^{-11}$ \\
Pyrex & 2.4~$\cdot$~10$^{-12}$ & 8.4~$\cdot$~10$^{-11}$ \\
Borosilicate Glass  &3.8~$\cdot$~10$^{-12}$ & 1.2~$\cdot$~10$^{-10}$ \\
Titanium  & 4.4~$\cdot$~10$^{-16}$ & 9.3~$\cdot$~10$^{-11}$ \\
Copper    & 0.0~$\cdot$~10$^{-11}$ & 9.5~$\cdot$~10$^{-12}$ \\
PE (C$_{2}$H$_{4}$) & 1.6~$\cdot$~10$^{-13}$ & 5.1~$\cdot$~10$^{-12}$ \\ 
PTFE (CF$_{2}$) & 7.1~$\cdot$~10$^{-12}$ & 7.7~$\cdot$~10$^{-10}$ \\

\end{tabular}

\end{center}
\end{table}

\setlength\tabcolsep{5pt}
\begin{table}
\caption{Radiogenic neutron yield (n$\cdot$s$^{-1}$$\cdot$cm$^{-3}$) for copper and polyethylene materials from $^{238}$U and $^{232}$Th decay chains.  Column (1)  and (2) refer to pure USD and SOURCES-4A calculation, respectively. Column (3) refers to SOURCES-4A calculation with USD ($\alpha$, n) cross section libraries. A ratio of the neutron yield is also provided: column (a) refers to the ratio of (2) over (1), whereas column (b) corresponds to the ratio (2)/(3)} 
\label{tab:cu-poly}
\begin{tabular}{ l  C{1.cm} C{1.4cm}  C{1.4cm} C{1.4cm}| C{0.9cm}C{0.9cm} } \\

&  & \multicolumn{3}{c}{\textbf{Neutron Yield  }} & &\\
&  & \multicolumn{3}{c}{\textbf{(10$^{-12}\cdot$~n$\cdot$s$^{-1}$$\cdot$cm$^{-3}$) }} & \multicolumn{2}{c}{\textbf{Ratio}}\\
\textbf{Material} & \textbf{Chain} & \textbf{ (1)} & \textbf{(2)} & \textbf{(3)} & \textbf{(a)} &  \textbf{(b)}\\ 
\hline
\hline
\multirow{2}{*}{Copper} & $^{238}$U & 3.46 & 2.84 & 2.93 & 0.8 &1.0\\
& $^{232}$Th & 11.1 & 9.49 & 9.18 & 0.9 & 1.0\\
\cline{2-7}
\multirow{2}{*}{Polyethylene (C$_{2}$H$_{4}$)} & $^{238}$U & 9.56 & 12.6 & 16.4 & 1.3 & 0.8\\
& $^{232}$Th & 2.87 & 5.28& 5.97 &  1.8 & 0.9 \\

%

\end{tabular}

\end{table}%

\setlength\tabcolsep{10pt}
\begin{table}[]
\caption{Radiogenic neutron yield (n$\cdot$s$^{-1}$$\cdot$cm$^{-3}$) per material considering 1~ppb of $^{238}$U and $^{232}$Th. The  percentage difference is calculates as (SOURCES-4A - USD)/[(SOURCES-4A + USD)/2].} 
\label{tab:radneutron}
\begin{center}
\tabcolsep1pt\begin{tabular}{l C{1.9cm}C{2.9cm}C{2.9cm} | C{1.4cm}} \\
&	&	\multicolumn{2}{c}{\textbf{Neutron Yield }}		&		 \\	
&	&	\multicolumn{2}{c}{\textbf{(n$\cdot$s$^{-1}$$\cdot$cm$^{-3}$)}}		&		 \\				
\textbf{Material}	&	\textbf{Chain}	&	{\scriptsize SOURCES-4A}	&	USD	&		\textbf{Diff \%}	\\
\hline											
\hline	

\multirow{2}{*}{Cu}	&	$^{238}$U	&	2.84~10$^{-12}$	&	3.46~10$^{-12}$	&	20	\\
	&	$^{232}$Th	&	9.49~10$^{-12}$	&	1.11~10$^{-11}$	&		16	\\
\cline{2-5}	
\multirow{2}{*}{PE (CH$_{2}$)}	&	$^{238}$U	&	1.26~10$^{-11}$	&	9.56~10$^{-12}$	&	-27	\\
	&	$^{232}$Th	&	5.28~10$^{-12}$	&	2.87~10$^{-12}$	&	-59	\\
\cline{2-5}	
\multirow{2}{*}{Titanium}	&	$^{238}$U	&	1.04~10$^{-10}$	&	1.99~10$^{-10}$	&		-63	\\
	&	$^{232}$Th	&	9.29~10$^{-11}$	&	1.24~10$^{-10}$	&		-28	\\
\cline{2-5}											
\multirow{2}{*}{Stainless Steel}	&	$^{238}$U	&	3.10~10$^{-11}$	&	5.95~10$^{-11}$	&		-63	\\
	&	$^{232}$Th	&	4.05~10$^{-11}	$&	6.80~10$^{-11}$	&		-51	\\
\cline{2-5}											
\multirow{2}{*}{Pyrex}	&	$^{238}$U	&	2.30~10$^{-10}$		&	1.61~10$^{-10}	$	&		36	\\
	&	$^{232}$Th	&	8.66~10$^{-11}$	&	4.59~10$^{-11}$	&		61	\\
\cline{2-5}											
\multirow{2}{*}{Borosilicate Glass}	&	$^{238}$U	&	3.48~10$^{-10}	$	&	2.45~10$^{-10}$		&		35	\\
	&	$^{232}$Th	&	1.27~10$^{-10}	$	&	6.98~10$^{-11}$	&		58	\\
\cline{2-5}											
\multirow{2}{*}{PTFE (CF$_{2}$)}	&	$^{238}$U	&	1.81~10$^{-9}$	&	1.60~10$^{-9}$	&		12	\\
	&	$^{232}$Th	&	7.76~10$^{-10}	$	&	5.42~10$^{-10}$		&		36	\\


\end{tabular}
\end{center}
\end{table}%

\setlength\tabcolsep{10pt}
\begin{table}[]
\caption{The differences in nuclear recoil counts over threshold simulated for different origin and target materials with SOURCES 4A and USD initial neutron spectra and yields. The  percentage difference is calculated as (SOURCES4A-USD)/[(SOURCES4A+USD)/2]. The $\chi^2$ per degree of freedom is calculated just for the single recoil spectra shape and excludes the normalization to total neutron yield.} 
\label{tab:recoilcounts}
\begin{center}
\tabcolsep1pt\begin{tabular}{l C{1.2cm}C{1.7cm}C{1.7cm}C{1.2cm}} \\
\textbf{Materia/Target}	&	\textbf{Chain}	&	Recoils Diff\%	&	Singles Diff\%	&  $\chi^2$/NDF	\\
\hline											
\hline	

\multirow{2}{*}{Borosilicate/Ar}	&	$^{238}$U	&	23	&	34	&	1.24	\\
	&	$^{232}$Th	&	27	&	41	&	1.32	\\
\cline{2-5}											
\multirow{2}{*}{PTFE/Xe}	&	$^{238}$U	&	-2	&	11	&	1.89	\\
	&	$^{232}$Th	&	23	&	26	&	1.06	\\
\cline{2-5}											
\multirow{2}{*}{Cu/Ge}	&	$^{238}$U	&	-81	&	-58	&	152	\\
	&	$^{232}$Th	&	-16	&	-14	&	5.83	\\	
	
\end{tabular}
\end{center}
\end{table}%

\end{document}